\documentclass[twocolumn]{aastex62}
\usepackage{amssymb}
\usepackage[utf8]{inputenc}
\usepackage{xspace}
\usepackage{graphicx}
\usepackage{amsmath}
\usepackage{url}
\usepackage{graphicx}
\usepackage{hyperref}
\usepackage{verbatim}
\usepackage{xcolor}
\usepackage{amssymb}

\newcommand{\psra}{J0405+3347\xspace}
\newcommand{\psrb}{J0742+4110\xspace}
\newcommand{\psrc}{J1018$-$1523\xspace}
\newcommand{\psrd}{J1045$-$0436\xspace}
\newcommand{\psre}{J1122$-$3546\xspace}
\newcommand{\psrf}{J1221$-$0633\xspace}
\newcommand{\psrg}{J1317$-$0157\xspace}
\newcommand{\psrh}{J1742$-$0203\xspace}
\newcommand{\psri}{J2017$-$2737\xspace}
\newcommand{\psrj}{J2018$-$0414\xspace}
\newcommand{\psrk}{J2022+2534\xspace}
\newcommand{\psrl}{J2039$-$3616\xspace}

\newcommand{\dmu}{pc\,cm$^{-3}$\xspace}
\newcommand{\Msun}{M$_{\odot}$\xspace}
\newcommand{\Rsun}{R$_{\odot}$\xspace}
\newcommand{\us}{$\mu$s\xspace}
\newcommand{\ps}{$\dot{P}_{\rm S}$\xspace}
\newcommand{\pg}{$\dot{P}_{\rm G}$\xspace}
\newcommand{\pint}{$\dot{P}_{\rm int}$\xspace}

\newcommand{\phnd}{\phantom{$-$}}

\newcommand{\km}{\textrm{km}}
\newcommand{\kpc}{\textrm{kpc}}

\newcommand{\s}{\textrm{s}}

\newcommand{\yr}{\textrm{yr}}

\newcommand{\erg}{\textrm{erg}}
\newcommand{\gauss}{\textrm{G}}

\shorttitle{12 New Pulsar Timing Solutions} 
\shortauthors{Swiggum et al.}

\begin{document}

\title{The Green Bank North Celestial Cap Survey. VII. 12 New Pulsar Timing Solutions}

\author[0000-0002-1075-3837]{J.~K.~Swiggum}
\affiliation{Center for Gravitation, Cosmology, and Astrophysics,
  Department of Physics, University of Wisconsin-Milwaukee, PO Box
  413, Milwaukee, WI, 53201, USA}
\affiliation{Dept. of Physics, 730 High St., Lafayette College, Easton, PA 18042, USA}

\author[0000-0002-4795-697X]{Z.~Pleunis}
\affiliation{Dunlap Institute for Astronomy \& Astrophysics,
  University of Toronto, 50 St. George Street, Toronto, Ontario, Canada M5S 3H4}

\author[0000-0002-0430-6504]{E.~Parent}
\affiliation{Dept.~of Physics and McGill Space Institute, McGill Univ., Montreal, QC H3A 2T8, Canada}

\author[0000-0001-6295-2881]{D.~L.~Kaplan}
\affiliation{Center for Gravitation, Cosmology, and Astrophysics,
  Department of Physics, University of Wisconsin-Milwaukee, PO Box
  413, Milwaukee, WI, 53201, USA}
  
\author[0000-0001-7697-7422]{M.~A.~McLaughlin}
\affiliation{Dept. of Physics and Astronomy, West Virginia University, Morgantown, WV 26506}
\affiliation{Center for Gravitational Waves and Cosmology, West Virginia University, Chestnut Ridge Research Building, Morgantown, WV 26506, USA}

\author[0000-0001-9784-8670]{I.~H.~Stairs}
\affiliation{Dept. of Physics and Astronomy, University of British Columbia, 6224 Agricultural Road, Vancouver, BC V6T 1Z1 Canada}

\author[0000-0002-6730-3298]{R.~Spiewak}
\affiliation{Jodrell Bank Centre for Astrophysics, School of Physics and Astronomy, The University of Manchester, Manchester, M13 9PL, UK}

\author[0000-0001-5134-3925]{G.~Y.~Agazie}
\affiliation{Center for Gravitation, Cosmology, and Astrophysics,
  Department of Physics, University of Wisconsin-Milwaukee, PO Box
  413, Milwaukee, WI, 53201, USA}

\author[0000-0002-3426-7606]{P.~Chawla}
\affiliation{Dept.~of Physics and McGill Space Institute, McGill Univ., Montreal, QC H3A 2T8, Canada}

\author[0000-0002-2185-1790]{M.~E.~DeCesar}
\affiliation{George Mason University, Fairfax, VA 22030, resident at the U.S. Naval Research Laboratory, Washington, D.C. 20375, USA}

\author[0000-0001-8885-6388]{T.~Dolch}
\affiliation{Department of Physics, Hillsdale College, 33 E. College Street, Hillsdale, MI 49242, USA}
\affiliation{Eureka Scientific, 2452 Delmer Street, Suite 100, Oakland, CA 94602-3017, USA}

\author[0000-0001-5645-5336]{W.~Fiore}
\affiliation{Dept. of Physics and Astronomy, West Virginia University, Morgantown, WV 26506}
\affiliation{Center for Gravitational Waves and Cosmology, West Virginia University, Chestnut Ridge Research Building, Morgantown, WV 26506, USA}

\author[0000-0001-8384-5049]{E.~Fonseca}
\affiliation{Dept. of Physics and Astronomy, West Virginia University, Morgantown, WV 26506}
\affiliation{Center for Gravitational Waves and Cosmology, West Virginia University, Chestnut Ridge Research Building, Morgantown, WV 26506, USA}

\author[0000-0002-8811-8171]{A.~G.~Istrate}
\affiliation{Department of Astrophysics/IMAPP, Radboud University, PO Box 9010, 6500 GL Nijmegen, The Netherlands}

\author[0000-0001-9345-0307]{V.~M.~Kaspi}
\affiliation{Dept.~of Physics and McGill Space Institute, McGill Univ., Montreal, QC H3A 2T8, Canada}

\author[0000-0001-8864-7471]{V.~I.~Kondratiev}
\affiliation{ASTRON, the Netherlands Institute for Radio Astronomy, Oude Hoogeveensedijk 4, 7991 PD Dwingeloo, The Netherlands}

\author[0000-0001-8503-6958]{J.~van Leeuwen}
\affiliation{ASTRON, the Netherlands Institute for Radio Astronomy, Oude Hoogeveensedijk 4, 7991 PD Dwingeloo, The Netherlands}

\author[0000-0002-2034-2986]{L.~Levin}
\affiliation{Jodrell Bank Centre for Astrophysics, School of Physics and Astronomy, The University of Manchester, Manchester, M13 9PL, UK}

\author[0000-0002-2972-522X]{E.~F.~Lewis}
\affiliation{Dept. of Physics and Astronomy, West Virginia University, Morgantown, WV 26506}
\affiliation{Center for Gravitational Waves and Cosmology, West Virginia University, Chestnut Ridge Research Building, Morgantown, WV 26506, USA}

\author[0000-0001-5229-7430]{R.~S.~Lynch}
\affiliation{Green Bank Observatory, P.O. Box 2, Green Bank, WV 24494, USA}

\author[0000-0001-5481-7559]{A.~E.~McEwen}
\affiliation{Center for Gravitation, Cosmology, and Astrophysics,
  Department of Physics, University of Wisconsin-Milwaukee, PO Box
  413, Milwaukee, WI, 53201, USA}

\author[0000-0002-4187-4981]{H.~Al Noori}
\affiliation{Dept. of Physics, University of California, Santa Barbara, CA 93106, USA}

\author[0000-0001-5799-9714]{S.~M.~Ransom}
\affiliation{National Radio Astronomy Observatory, 520 Edgemont Rd., Charlottesville, VA 22903, USA}

\author[0000-0002-7778-2990]{X.~Siemens}
\affiliation{Dept. of Physics, Oregon State University, Corvallis, OR 97331, USA}

\author[0000-0002-9507-6985]{M.~Surnis}
\affiliation{Department of Physics, IISER Bhopal, Bhauri Bypass Road, Bhopal 462066, India}

\correspondingauthor{J.~K.~Swiggum}
\email{swiggumj@lafayette.edu}

\begin{abstract}
We present timing solutions for 12 pulsars discovered in the Green Bank North Celestial Cap (GBNCC) 350\,MHz pulsar survey, including six millisecond pulsars (MSPs), a double neutron star (DNS) system, and a pulsar orbiting a massive white dwarf companion. Timing solutions presented here include 350 and 820\,MHz Green Bank Telescope data from initial confirmation and follow-up as well as a dedicated timing campaign spanning one year. PSR~\psre is an isolated MSP, PSRs~\psrf and \psrg are MSPs in black widow systems and regularly exhibit eclipses, and PSRs~\psrk and \psrl are MSPs that can be timed with high precision and have been included in pulsar timing array experiments seeking to detect low-frequency gravitational waves. PSRs \psrf and \psrl have {\it Fermi} Large Area Telescope $\gamma$-ray counterparts and also exhibit significant $\gamma$-ray pulsations. We measure proper motion for three of the MSPs in this sample and estimate their space velocities, which are typical compared to those of other MSPs. We have detected the advance of periastron for PSR~\psrc and therefore measure the total mass of the double neutron star system, $m_{\rm tot}=2.3\pm0.3$\,\Msun. Long-term pulsar timing with data spanning more than one year is critical for classifying recycled pulsars, carrying out detailed astrometry studies, and shedding light on the wealth of information in these systems post-discovery.
\end{abstract}


\section{Introduction}
The Green Bank North Celestial Cap (GBNCC) pulsar survey began in 2009 and is largely complete, having discovered 194 pulsars so far. Using the 100\,m Green Bank Telescope (GBT), the survey has covered the full sky accessible to the Green Bank Observatory, all declinations $\delta>-40^{\circ}$. Operating at a relatively low center frequency of 350\,MHz, individual beams are $36\arcmin$ across and the large survey region (85\% of the celestial sphere) can be covered efficiently with $\approx125$,000 overlapping pointings. With overhead, this comes out to $\approx5$,500\,hours of scheduled telescope time. Only a few observations remain, re-observing pointings being significantly affected by radio frequency interference (RFI).

For each 120-s sky pointing, data are collected in search mode with 4096 frequency channels spanning 100\,MHz of bandwidth centered at 350\,MHz. Total intensities are sampled every 81.92\,\us. Data are transferred to McGill University and processed using large allocations on Compute Canada supercomputers.
Searches for both periodic and transient signals are carried out at a range of trial dispersion measures (DMs) with a pipeline based on the {\tt PRESTO}\footnote{\url{https://www.cv.nrao.edu/~sransom/presto/}} software package \citep{rem+02}. 

A full description of the survey and initial sensitivity projections can be found in \cite{slr+14}. Timing solutions for GBNCC discoveries are included there, in \cite{ksr+12}, \cite{kkl+15}, and more recently, \cite{kmk+18}, \cite{lsk+18}, \cite{acd+19}, and \cite{amm+21}. The first fast radio burst discovery (FRB20200125A) is described in \cite{pck+20}, and \cite{mss+20} provides a census of GBNCC discoveries at the time of publication and detailed survey sensitivity analysis. Published standard profiles, pulse times of arrival, and timing models from most of these previous studies are publicly available on GitHub,\footnote{\url{https://github.com/GBNCC/data}} linked from the GBNCC discoveries page.\footnote{\url{http://astro.phys.wvu.edu/GBNCC}}

The primary science goal of the GBNCC pulsar survey is discovering millisecond pulsars (MSPs), a class of old neutron stars spun up through mass transfer from a donor companion \citep{alpar+82}. High-precision MSP discoveries are critical for the detection of a stochastic nanohertz gravitational wave (GW) background. Such a background signal would likely come from coalescing super-massive black holes \citep{jb+03}, relic cosmological gravitational waves \citep[e.g.][]{grishchuk+05}, and/or cosmic strings \citep{maggiore+00}. Nanohertz GW detection efforts with pulsar timing arrays (PTAs) span the globe; the US--Canada effort, the North American Nanohertz Observatory for Gravitational waves (NANOGrav), most recently published a 12.5-yr data release \citep{NG_12.5nb,NG_12.5wb} where the first hints of a GW background may be present \citep{NG_12.5gwb}. The latest combined International Pulsar Timing Array (IPTA) data release \citep[DR2;][]{DR2}, comprised of PTA data sets from groups in Europe, Australia, and North America, shows similar hints of the GW background \citep{DR2gwb}. The most effective way to increase PTA sensitivity to the nanohertz GW background is by adding MSPs to the array \citep{sej+13}. NANOGrav aims to add four MSPs to its array each year, and thus relies heavily on pulsar surveys like GBNCC to provide these new sources. 

The GBNCC pulsar survey also aims to find exotic new binary systems that push the boundaries of our understanding in various areas. Double neutron star (DNS) systems can place constraints on NS kick distributions \citep{tkf+17,vns+18}, provide laboratories for testing theories of gravity \citep[for a review, see][]{will+14}, and inform NS merger rates \citep{bdp+03,kkl+04,cbk+18,mb+22}. Eclipsing binaries offer opportunities to probe material surrounding the companion in/around the eclipse. In some cases, pulsar binaries can also constrain the equation of state of supranuclear matter via NS mass measurements \citep{cfr+20}. All of these systems bring into focus the wide variety of evolutionary scenarios and offer possible explanations for open questions (e.g., the origin of isolated MSPs), and in many cases, even richer information can be gleaned from multi-frequency follow-up \citep{skm+17}. In this study, we have specifically targeted new discoveries with spin periods $<200$\,ms for timing follow-up to identify recycled pulsars, distinguish between isolated and binary systems, and start tackling some of these broader science goals. 

Section \ref{sec:oa} describes the confirmation and timing follow-up for 12 discoveries, including flux density and spectral index measurements based on observations at 350 and 820\,MHz. Timing model parameters and their values are presented in Section~\ref{sec:results}, including further analysis for three MSP systems where proper motions were detected. Section \ref{sec:disc} describes the process we used to search for $\gamma$-ray counterparts (and pulsations, where appropriate), as well as individual source classifications and interesting features based on timing models. We conclude in Section \ref{sec:conclusion} and outline some future work that is underway.

\begin{figure*}[t]
\centering
\includegraphics[width=0.9\textwidth]{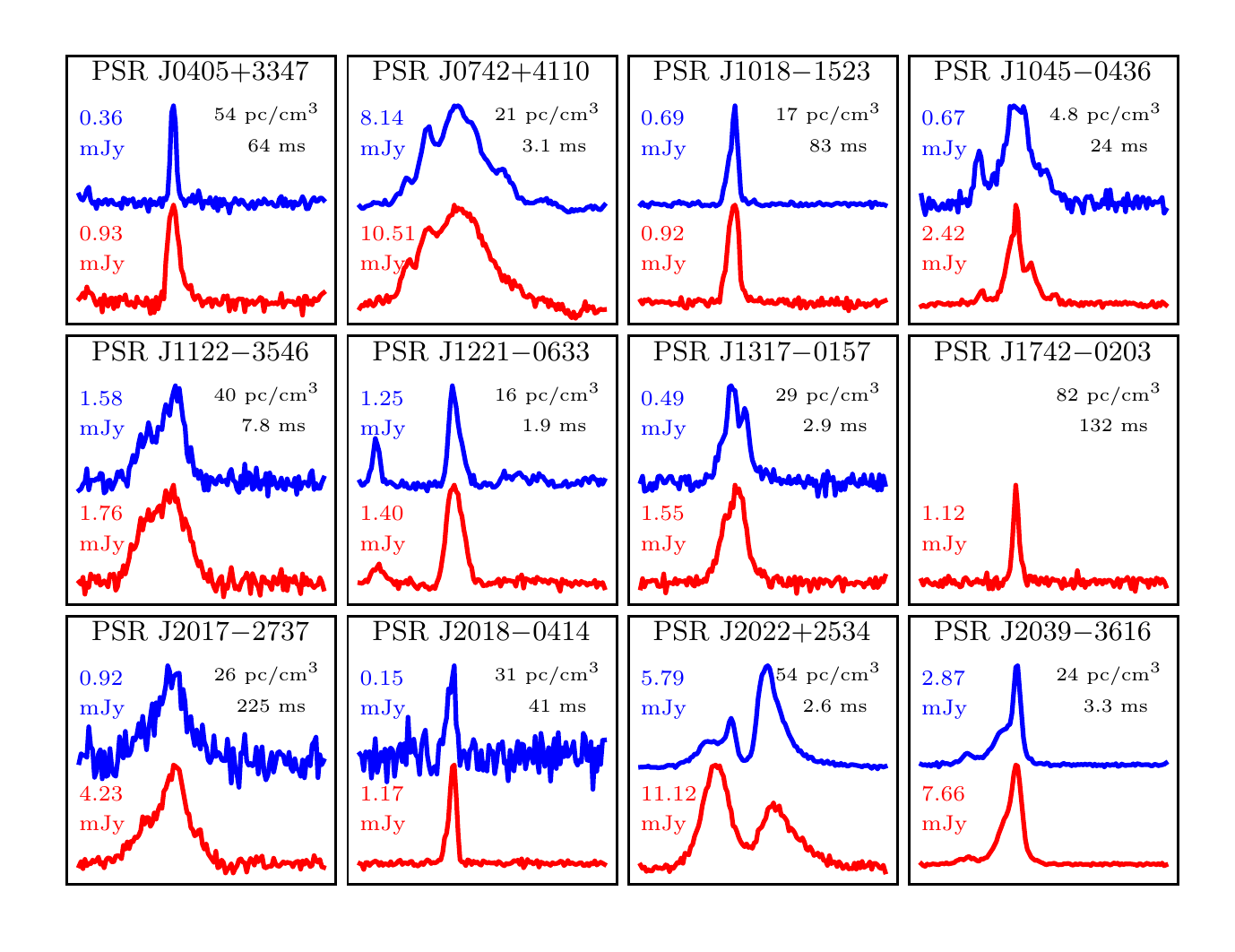}
\centering
\caption{Summed pulse profiles depicting intensity over a full rotation. Profiles have been scaled by their respective maximum intensities to make it easier to compare their shapes. They were generated using all 350 MHz (red; bottom of each panel) and 820 MHz (blue; top) data available, with corresponding flux densities in the same colors listed to the left in mJy. To the right of each profile, the pulsar's DM is given in pc\,cm$^{-3}$ and spin period in ms. PSR~\psrh was never detected at 820 MHz.}
\label{fig:profs}
\end{figure*}

\section{Pulsar Timing Observations \& Analysis}\label{sec:oa}
\subsection{Observations}\label{sec:obs}
The twelve discoveries described in this paper were initially flagged as periodicity candidates, then confirmed with GBT scans at 350\,MHz. All were used regularly as test sources during survey observations and folded in real time to monitor the RFI environment and survey data quality. These data served a dual purpose since they were also included in our timing analysis (Section \ref{sec:ta}). Test scans conducted at 350\,MHz used 81.92\,\us time resolution and 4096 channels across 100\,MHz of bandwidth.

Following confirmation scans at 350\,MHz, pulsar positions were improved using an on-the-fly mapping technique described in \cite{sg+18}, resulting in position uncertainties of $\approx1-3$\arcmin. Improved localization ensures that the pulsar is closer to the telescope's boresight in observations that follow; it provides additional flexibility in choice of observing frequency (since telescope beam size is inversely proportional to the chosen center frequency), ensures higher signal-to-noise detections and thus, more efficient follow-up, and facilitates the process of finding an initial timing solution.

Afterwards, timing data were collected using the GBT (project code 17B$-$285; PI: J.\ Swiggum) at 820\,MHz with 2048 channels across 200\,MHz bandwidth and 40.96\,\us time resolution. For PSR~\psrb, timing data were included from a previous GBT timing campaign using the same setup (project code 15A$-$376; PI: L.\ Levin). Each pulsar was observed with a monthly cadence over a full year; high-cadence sessions were also included to observe each pulsar $4-5$ times over one week to establish initial phase connection and aid in solving binary parameters, where necessary.

\begin{deluxetable*}{lrclcrclcc}
  \centering
  \tabletypesize{\footnotesize}
  \tablewidth{0pt}
  \tablecolumns{9}
  \tablecaption{350 \& 820 MHz Flux Densities, Spectral Index Measurements}
  \tablehead{
        \colhead{}                  &
	\multicolumn{3}{c}{350 MHz} &
	\colhead{}                  &
	\multicolumn{3}{c}{820 MHz} &
	\colhead{} \\
	\cline{2-4} \cline{6-8}
    \colhead{PSR}                      &
    \colhead{$t_{\rm int}$}            &
    \colhead{$\delta$}                 &
    \colhead{$S_{350}$}                &
    \colhead{}                         &
    \colhead{$t_{\rm int}$}            &
    \colhead{$\delta$}                 &
    \colhead{$S_{820}$}                &
    \colhead{$\alpha$} \\
    \colhead{}                         &
    \colhead{(s)}                      &
    \colhead{}                         &
    \colhead{(mJy)}                    &
    \colhead{}                         &
    \colhead{(s)}                      &
    \colhead{}                         &
    \colhead{(mJy)}                    &
    \colhead{}                            
  }
  \startdata
J0405+3347 & 3309.8 & 0.17 & \phn0.93(13) & & 584.0 & 0.08 & 0.36(7) & $-$1.1(3) \\
J0742+4110 & 5906.3 & 0.81 & 10.5(1.5) & & 5480.4 & 0.86 & 8.1(1.8) & $-$0.3(3) \\
J1018$-$1523 & 2583.0 & 0.12 & \phn0.92(14) & & 3686.3 & 0.16 & 0.69(14) & $-$0.3(3) \\
J1045$-$0436 & 14112.3 & 0.45 & \phn2.4(4) & & 4944.2 & 0.41 & 0.67(13) & $-$1.5(3) \\
J1122$-$3546 & 4015.8 & 0.38 & \phn1.8(3) & & 1188.0 & 0.31 & 1.6(8) & $-$0.1(7) \\
J1221$-$0633 & 10225.4 & 0.32 & \phn1.4(2) & & 5841.1 & 0.59 & 1.2(2) & $-$0.1(3) \\
J1317$-$0157 & 9557.0 & 0.30 & \phn1.6(2) & & 4309.6 & 0.28 & 0.49(9) & $-$1.4(3) \\
J1742$-$0203 & 4539.9 & 0.11 & \phn1.12(15) & & -- & (0.06) & $<0.2$ & $<-1.5$ \\
J2017$-$2737 & 3935.3 & 0.47 & \phn4.2(6) & & 564.0 & 0.38 & 0.92(2) & $-$1.8(3) \\
J2018$-$0414 & 8350.4 & 0.12 & \phn1.17(16) & & 564.0 & 0.09 & 0.15(3) & $-$2.4(3) \\
J2022+2534 & 6534.4 & 0.66 & 11.1(1.4)\phn & & 3888.0 & 0.66 & 6(1) & $-$0.8(3) \\
J2039$-$3616 & 7637.0 & 0.45 & \phn7.7(1.1) & & 4068.6 & 0.48 & 2.9(5) & $-$1.2(3) \\
  \enddata
  \tablecomments{Total integration time ($t_{\rm int}$) used to generate profiles, and measured duty cycles ($\delta$), flux densities ($S_{350}$/$S_{820}$), and spectral indices ($\alpha$) are listed for pulsars at each observing frequency included in our analysis. Since \psrh was not detected at 820 MHz, we place limits on $S_{820}$ and $\alpha$ for this pulsar, assuming a typical duty cycle, $\delta=0.06$.}
\end{deluxetable*}
\label{tab:flux}

\subsection{Measured Flux Densities: $S_{350}$ \& $S_{820}$}\label{sec:profs}
Since data collection for this study was sometimes opportunistic and/or coherent timing solutions for sources included were not initially available (see Section \ref{sec:obs}), observations were predominantly conducted in search mode and time was not spent on polarization/flux density calibration. Therefore, we estimate 350 and 820 MHz flux densities ($S_{350}$ and $S_{820}$) for each source using summed, total intensity pulse profiles (see Figure \ref{fig:profs}) and the radiometer equation as follows.

As in \cite{lk+04}, flux densities at respective observing frequencies, $S_\nu$, are computed here using the radiometer equation,
\begin{equation}
S_\nu = \beta\,\frac{({\rm S/N})\,T_{\rm sys}}{G(\theta)\,\sqrt{n_{\rm p}\,t_{\rm int}\,\Delta f}}\,\sqrt{\frac{\delta}{1-\delta}},
\label{eqn:radiometer}
\end{equation}
where signal-to-noise (S/N) is measured from summed pulse profiles shown in Figure \ref{fig:profs}, using the same technique as described in \cite{mss+20}. System temperature ($T_{\rm sys}$) is the sum of sky temperature ($T_{\rm sky}$) and receiver temperature ($T_{\rm rec}$). We use {\tt PyGDSM}\footnote{\url{https://github.com/telegraphic/pygdsm}} to get $T_{\rm sky}$, including the contribution from the cosmic microwave background, at each source position and observing frequency based on \cite{ztd+17}. At 350/820\,MHz, $T_{\rm rec}=23/22$\,K, respectively (see Figure 3 in the GBO Proposer's Guide\footnote{\url{https://www.gb.nrao.edu/scienceDocs/GBTpg.pdf}}). Duty cycle ($\delta$) here is the fraction of the integrated profile where the pulsar's signal is present, $\delta=n_{\rm on}/n_{\rm bin}$. Degradation due to digitization is reflected by $\beta=1.3$, number of summed polarizations is $n_{\rm p}=2$, effective bandwidth is $\Delta f=70/175$\,MHz at 350/820\,MHz, respectively (accounting for common RFI zapping), and the GBT's gain along the boresight, $G(0)=2$\,K/Jy. Since timing positions were not measured until relatively late in our follow-up programs, observing catalog positions -- even after improvements -- could be offset by several arcminutes. These offsets translate to some amount of degradation in the effective gain. Gaussian functions with full-width-half-maxima (FWHM) equal to those of the 350/820\,MHz beams (${\rm FWHM}=36\arcmin/15\arcmin$ respectively) provide good approximations of degradation as a function of position offset. For our flux density analysis, we generate integrated pulse profiles using observations within $\theta=7\arcmin/3\arcmin$ at 350/820\,MHz, respectively, which translates to  a 10\% degradation in gain. In several cases, tolerating larger offsets was required to integrate a sufficient number of observations, and larger degradation factors were applied when calculating $S_{820}$ for PSRs~\psrb, \psre, and \psri.  

Table \ref{tab:flux} lists total integration times ($t_{\rm int}$) for individual sources for each observing frequency, as well as  measured duty cycles ($\delta$), and resulting flux densities and spectral indices ($\alpha$, where $S_\nu\propto\nu^\alpha$). To estimate uncertainties, we use standard error propagation assuming $\sigma_{\Delta f}=10$\,MHz (due to transient sources of RFI, effective bandwidth can vary), $\sigma_{T_{\rm sys}}=5$\,K, and $\sigma_G=0.1-0.5$\,K/Jy, depending on typical observing position offsets.

To check our measurements for consistency, we looked at the literature and other catalogs for matching detections and flux density measurements. In a census of MSP flux densities with MeerKAT, \cite{sbm+22} found $S_{1400}=0.50\pm0.04$\,mJy for PSR~\psrl, and a spectral index of $-2.0\pm0.4$. These values, scaled to 350\,MHz, are completely consistent with $S_{350}$ reported here, but our $S_{820}$ measurement (and therefore spectral index) is only consistent at the $2-3\sigma$ level. PSR~\psrk was detected in the Rapid ASKAP Continuum Survey, RACS-low \citep[888\,MHz;][]{RACSdr1} with $S_{888}=3.6\pm0.3$\,mJy, which is consistent with our measurement at the $\approx2\sigma$ level. Other catalogs such as TIFR GMRT Sky Survey (TGSS) and LOFAR Two-meter Sky Survey (LoTSS) did not have any unidentified radio sources corresponding to those in this sample. Finally, we compared $S_{350}$ measurements here with those presented in \cite{mss+20} and find broad consistency at the $\approx1-2\sigma$ level (except for PSRs~\psrb and \psrl, whose values here are about five times higher). At 350\,MHz both of these pulsars have estimated scintillation timescales ($\approx250$-s and $\approx120$-s, respectively) near the length of a GBNCC survey observation time (120-s), so it is plausible they were scintillated down during their discovery scans.

Due to systematic uncertainties often present in estimating flux densities using the radiometer equation, measured values can be discrepant by factors of two or more. Taking this into account, our measurements are in reasonable agreement with those from other radio surveys and previous studies.

Finally, we note some surprise at the fact that none of the 350\,MHz summed profiles in Figure \ref{fig:profs} exhibit significant scatter broadening compared to their 820\,MHz counterparts. Electron density models (NE2001/YMW16) predict scattering timescales at the level of $5-30$\% of a rotation in most cases, but $>50$\% for \psrk. Many of these sources are well off the Galactic plane (see Table \ref{tab:astrometric}) where electron density models tend to have higher uncertainties, but this may also be a selection effect. Sources that exhibit less scattering are more likely to be detected in the 350\,MHz GBNCC pulsar survey.

\subsection{Detections \& Preliminary Binary Parameters}\label{sec:binprepar}
Before timing solutions were available, periodicity searches were carried out using dedispersed timeseries from each epoch since many sources were known to be in binary systems and therefore their apparent spin periods would change between sessions. First, RFI was masked automatically using {\tt rfifind} and the known DM was applied to produce topocentric and barycentric timeseries with {\tt prepdata}. Periodicity candidates were generated with {\tt accelsearch}.
After finding a candidate period close to the discovery value, the raw data were folded using {\tt prepfold} and the candidate periodicity, refined by allowing a fine search in period and period derivative.

\begin{figure}
\centering
\includegraphics[width=0.45\textwidth]{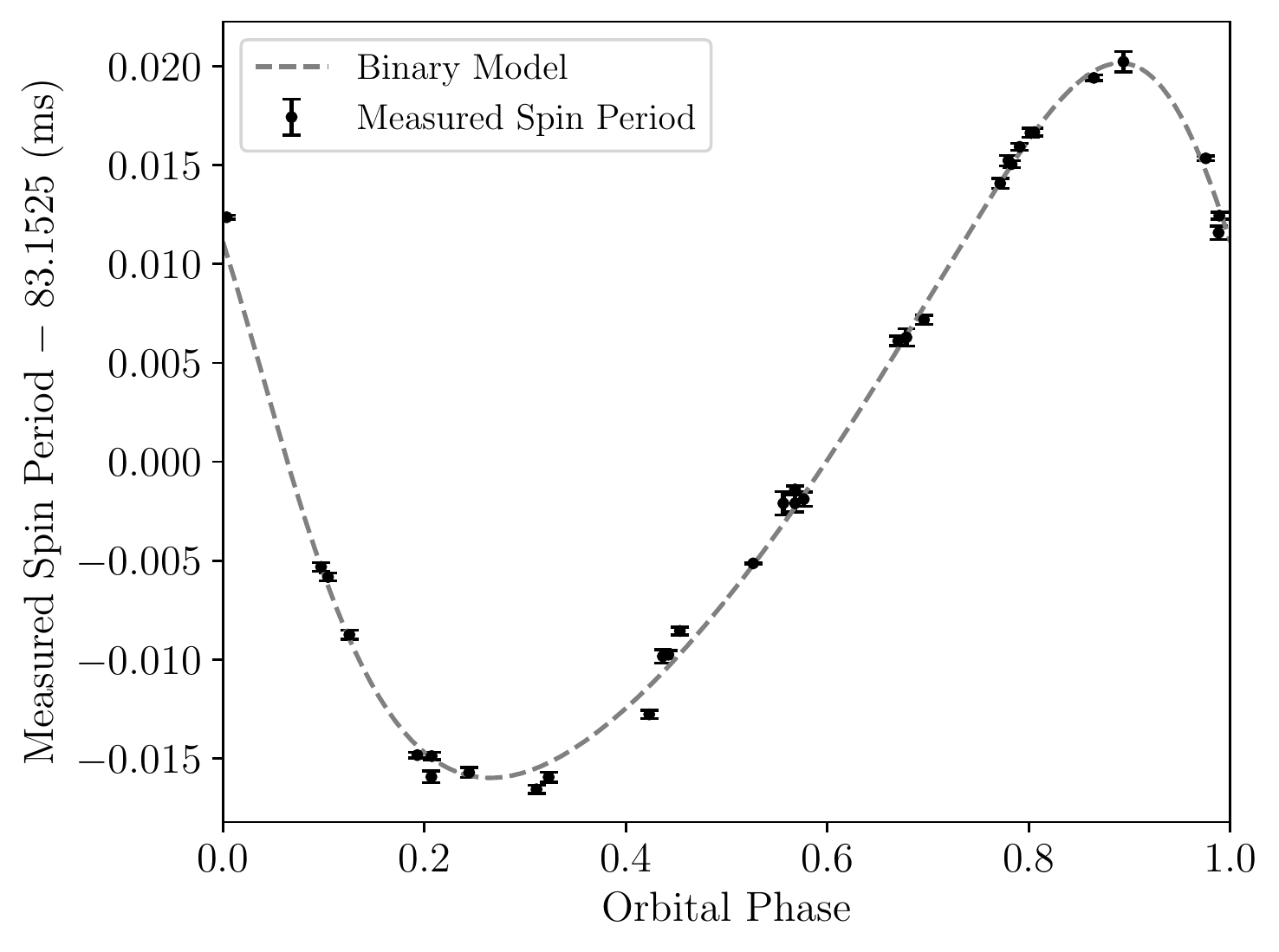}
\centering
\caption{Spin period measurements and 1-$\sigma$ uncertainties for PSR~\psrc plotted as a function of orbital phase. The gray dashed line represents expected apparent spin period changes as a function of orbital phase (mean anomaly), given a preliminary set of orbital parameters similar to those reported in Table \ref{tab:dns}.}
\label{fig:prebin}
\end{figure}

Binary systems were identified by their time-variable barycentric spin periods, which were compiled and analyzed for each source using the ``roughness" method described in \cite{bn+08}. Roughness,
\begin{equation}
\mathcal{R}=\sum_{i=1}^{n-1}[P_{\rm obs}(i)-P_{\rm obs}(i+1)]^2,
\label{eqn:rough}
\end{equation}
where $P_{\rm obs}$ represents a set of observed spin periods, sorted by their orbital phase using corresponding observation epochs and trial orbital period values. Roughness was calculated for many trial orbital periods and minimized to provide a reasonable guess for the best initial value. A full set of preliminary binary parameters followed for each binary system by devising a rough model that matched the shape of measured spin period versus orbital phase ($\phi_{\rm orb}$; see Figure \ref{fig:prebin}).

\begin{deluxetable*}{llrccccc}
  \tabletypesize{\footnotesize}
  \tablewidth{\textwidth}
  \tablecaption{Rotational and Timing Parameters of GBNCC Pulsars}
  \tablecolumns{8}
  \tablehead{
    \colhead{PSR} & \colhead{$\nu$} & 
    \colhead{$\dot{\nu}$} &
    \colhead{Epoch} & \colhead{Data Span} & 
    \colhead{RMS Residual} & \colhead{$N_{\rm TOA}$} & 
    \colhead{EFAC} \\ 
    \colhead{} & \colhead{(Hz)} & 
    \colhead{(Hz\,s$^{-1}$)} & 
    \colhead{(MJD)} & \colhead{(MJD)} & \colhead{(\us)} & \colhead{}}
  \startdata
  J0405$+$3347 & \phn15.6362508979(4) & $-$4.0(3)$\times$10$^{-17}$ & 57824 & 57445--58201 & 174.6 & 108 & 1.0597 \\
  J0742$+$4110 & 318.55889711523(3) & $-$6.791(4)$\times$10$^{-16}$ & 57169 & 56044--58294 & \phn12.1 & 222 & 1.3308 \\
  J1018$-$1523 & \phn12.02609153542(3) & $-$1.58(8)$\times$10$^{-17}$ & 57998 & 57542--58452 & \phn65.8 & 425 & 1.1679 \\
  J1045$-$0436 & \phn41.58433482255(5) & $-$1.36(1)$\times$10$^{-16}$ & 57891 & 57478--58304 & \phn45.5 & 153 & 1.3027 \\
  J1122$-$3546 & 127.5824382850(3) & $-$2.48(9)$\times$10$^{-16}$ & 57858 & 57449--58266 & 120.4 & 250 & 1.2736 \\
  J1221$-$0633 & 516.918832068(1) & $-$1.42(2)$\times$10$^{-15}$ & 58105 & 57906--58304 & \phn\phn4.4 & 264 & 1.2066 \\
  J1317$-$0157 & 343.850032475(2) & $-$6.5(4)$\times$10$^{-16}$ & 58107 & 57909--58304 & \phn23.5 & 164 & 1.2961 \\
  J1742$-$0203 & \phn\phn7.59822533(3) & $-$8.6(7)$\times$10$^{-15}$ & 58014 & 57909--58118 & 285.6 & \phn45 & 1.0817 \\
  J2017$-$2737 & \phn\phn4.4538744(1) & $-$1.21(1)$\times$10$^{-13}$ & 58041 & 57906--58174 & 1802.8\phn & \phn47 & 1.6155 \\
  J2018$-$0414 & \phn24.623136942(1) & $-$4(2)$\times$10$^{-17}$ & 58105 & 57906--58303 & \phn73.0 & 136 & 1.0262 \\
  J2022$+$2534 & 377.93812391457(8) & $-$8.80(2)$\times$10$^{-16}$ & 57919 & 57535--58303 & \phn10.0 & 744 & 1.0354 \\
  J2039$-$3616 & 305.33963750348(3) & $-$7.845(8)$\times$10$^{-16}$ & 57920 & 57537--58303 & \phn\phn4.3 & 375 & 1.0349 \\
  \enddata
  \tablecomments{All timing
  models use the DE430 Solar system ephemeris and are referenced to
  the TT(BIPM) time standard.  Values in parentheses are the
  $1$-$\sigma$ uncertainty in the last digit as reported by
  \texttt{TEMPO}. Multiplicative error factors (EFACs) listed here were applied to TOA uncertainties, forcing $\chi^2_{\rm red}=1$.}
\end{deluxetable*}
\label{tab:spin}
\begin{deluxetable*}{llllccrr}
  \tabletypesize{\footnotesize}
  \tablewidth{\textwidth}
  \tablecaption{Coordinates and DMs of GBNCC Pulsars}
  \tablecolumns{8}
  \tablehead{\colhead{PSR} & \multicolumn{3}{c}{Measured} & 
             \multicolumn{4}{c}{Derived} \\ 
             \colhead{} & \colhead{$\lambda$ ($\arcdeg$)} & 
             \colhead{$\beta$ ($\arcdeg$)} &
             \colhead{DM (\dmu)} & 
             \colhead{$\alpha$ (J2000)} & \colhead{$\delta$ (J2000)} & 
             \colhead{$\ell$ ($\arcdeg$)} & \colhead{$b$ ($\arcdeg$)}}
  \startdata
J0405+3347 & \phn65.908034(6) & \phnd12.70753(3) & 53.596(3) & $04^{\rm h}\, 05^{\rm m}\, 29\, \fs57$ & $+33\arcdeg\, 47\arcmin\, 00\, \farcs3$ & 162.78 & $-$13.68 \\
J0742+4110 & 110.1469948(7) & \phnd19.502018(1) & 20.8135(2) & $07^{\rm h}\, 42^{\rm m}\, 12\, \fs19$ & $+41\arcdeg\, 10\arcmin\, 14\, \farcs9$ & 178.13 & 26.57 \\
J1018$-$1523 & 162.497241(2) & $-$24.093012(8) & 17.158(2) & $10^{\rm h}\, 18^{\rm m}\, 12\, \fs72$ & $-15\arcdeg\, 23\arcmin\, 10\, \farcs2$ & 257.13 & 33.53 \\
J1045$-$0436 & 164.7121434(8) & $-$11.510665(4) & \phn4.8175(9) & $10^{\rm h}\, 45^{\rm m}\, 57\, \fs92$ & $-04\arcdeg\, 36\arcmin\, 23\, \farcs4$ & 254.46 & 46.12 \\
J1122$-$3546 & 187.868380(2) & $-$36.102612(4) & 39.5868(7) & $11^{\rm h}\, 22^{\rm m}\, 17\, \fs24$ & $-35\arcdeg\, 46\arcmin\, 31\, \farcs2$ & 283.30 & 23.67 \\
J1221$-$0633 & 187.5164776(2) & \phn$-$3.900195(2) & 16.43241(6) & $12^{\rm h}\, 21^{\rm m}\, 24\, \fs76$ & $-06\arcdeg\, 33\arcmin\, 51\, \farcs7$ & 289.68 & 55.53 \\
J1317$-$0157 & 198.6676339(8) & \phnd\phn5.786217(9) & 29.4008(2) & $13^{\rm h}\, 17^{\rm m}\, 40\, \fs45$ & $-01\arcdeg\, 57\arcmin\, 30\, \farcs1$ & 316.23 & 60.23 \\
J1742$-$0203 & 265.2820(4) & \phnd21.3053(2) & 81.82 & $17^{\rm h}\, 42^{\rm m}\, 24\, \fs53$ & $-02\arcdeg\, 03\arcmin\, 43\, \farcs2$ & 22.99 & 14.33 \\
J2017$-$2737 & 300.243(2) & \phn$-$7.600(4) & 25.82(5) & $20^{\rm h}\, 17^{\rm m}\, 01\, \fs63$ & $-27\arcdeg\, 30\arcmin\, 49\, \farcs2$ & 15.23 & $-$29.91 \\
J2018$-$0414 & 305.834964(6) & \phnd15.00891(1) & 30.914(1) & $20^{\rm h}\, 18^{\rm m}\, 10\, \fs41$ & $-04\arcdeg\, 14\arcmin\, 12\, \farcs7$ & 39.23 & $-$21.20 \\
J2022+2534 & 316.3805463(2) & \phnd43.4494254(2) & 53.6623(1) & $20^{\rm h}\, 22^{\rm m}\, 33\, \fs26$ & $+25\arcdeg\, 34\arcmin\, 42\, \farcs5$ & 66.10 & $-$6.54 \\
J2039$-$3616 & 302.72326687(7) & $-$17.2460168(3) & 23.96332(7) & $20^{\rm h}\, 39^{\rm m}\, 16\, \fs58$ & $-36\arcdeg\, 16\arcmin\, 17\, \farcs2$ & 6.33 & $-$36.52 \\
  \enddata
  \tablecomments{Ecliptic coordinates use the IERS2010 value of the
  obliquity of the ecliptic referenced to J2000 \citep{cwc03}. Values in parentheses are the $1$-$\sigma$ uncertainty in the last digit as reported by \texttt{TEMPO}. In some cases, the reported precision goes beyond month-year-timescale changes in DM that might be expected due to ISM effects \citep[e.g. see][]{jml+17}, but modeling those changes goes beyond the scope of this work.}
\end{deluxetable*}
\label{tab:astrometric}

\subsection{Timing Analysis}\label{sec:ta}
Before timing ephemerides were available for discoveries, individual scans were processed as described in Section \ref{sec:binprepar} and three times-of-arrival (TOAs) were generated per $5-10$\,minute observation with {\tt get\_TOAs.py} from {\tt PRESTO}. In order to accurately determine arrival times, a standard profile is cross-correlated with the observed signal in the Fourier domain \citep{taylor+92}. In this initial stage, a standard profile was generated for each pulsar with {\tt pygaussfit.py}, fitting Gaussian components to the highest signal-to-noise (S/N) profile available. Due to frequency-dependent profile evolution, different standard profiles were used for calculating TOAs at 350 and 820\,MHz as necessary. Three TOAs per epoch allowed fits for spin frequency on a per-epoch basis, facilitating phase connection over short time scales \--- initially days to weeks. These coherent timing solutions were then extended across the full data span using the {\tt TEMPO}\footnote{\url{http://tempo.sourceforge.net/}} pulsar timing software.

With coherent timing solutions in hand, individual scans were refolded and manipulated as follows using processing routines available in the {\tt PSRCHIVE}\footnote{\url{http://psrchive.sourceforge.net/}} software suite \citep{hsm+04}. First, RFI was carefully excised using {\tt pazi}, then scans were scrunched down to $3-5$ sub-integrations and $2-4$ sub-bands, signal-to-noise permitting, ensuring that the pulsar signal was detectable in each of these divisions. For each pulsar, detections were summed coherently at respective observing frequencies using {\tt psradd} \--- which uses ephemerides to phase-align observations from different epochs \--- to create averaged profiles (see Figure \ref{fig:profs} and Section \ref{sec:profs} for details regarding profile analysis). Noise-free standard profiles at 350 and 820\,MHz were generated by fitting Gaussian components to averaged profiles, and the two templates were aligned using {\tt pas}. Standard profiles were cross-correlated with folded, cleaned, and scrunched data to produce a final set of TOAs with {\tt pat}. Since standard profiles for respective observing bands were aligned as part of this process, we did not fit for any time offsets (``jumps") between corresponding TOAs, however, observing mode-dependent instrumental offsets were taken into account (e.g. 61.44\,\us instrumental delay between 350/820\,MHz data collected in incoherent mode).  Parameter fitting and refinement was conducted using {\tt TEMPO}, with the DE430 solar system ephemeris and TT(BIPM) time standard implemented therein. 

\section{results}\label{sec:results}
\subsection{Timing Model Fitting}\label{sec:fits}
\begin{figure*}[!htb]
\centering
\includegraphics[width=0.9\textwidth]{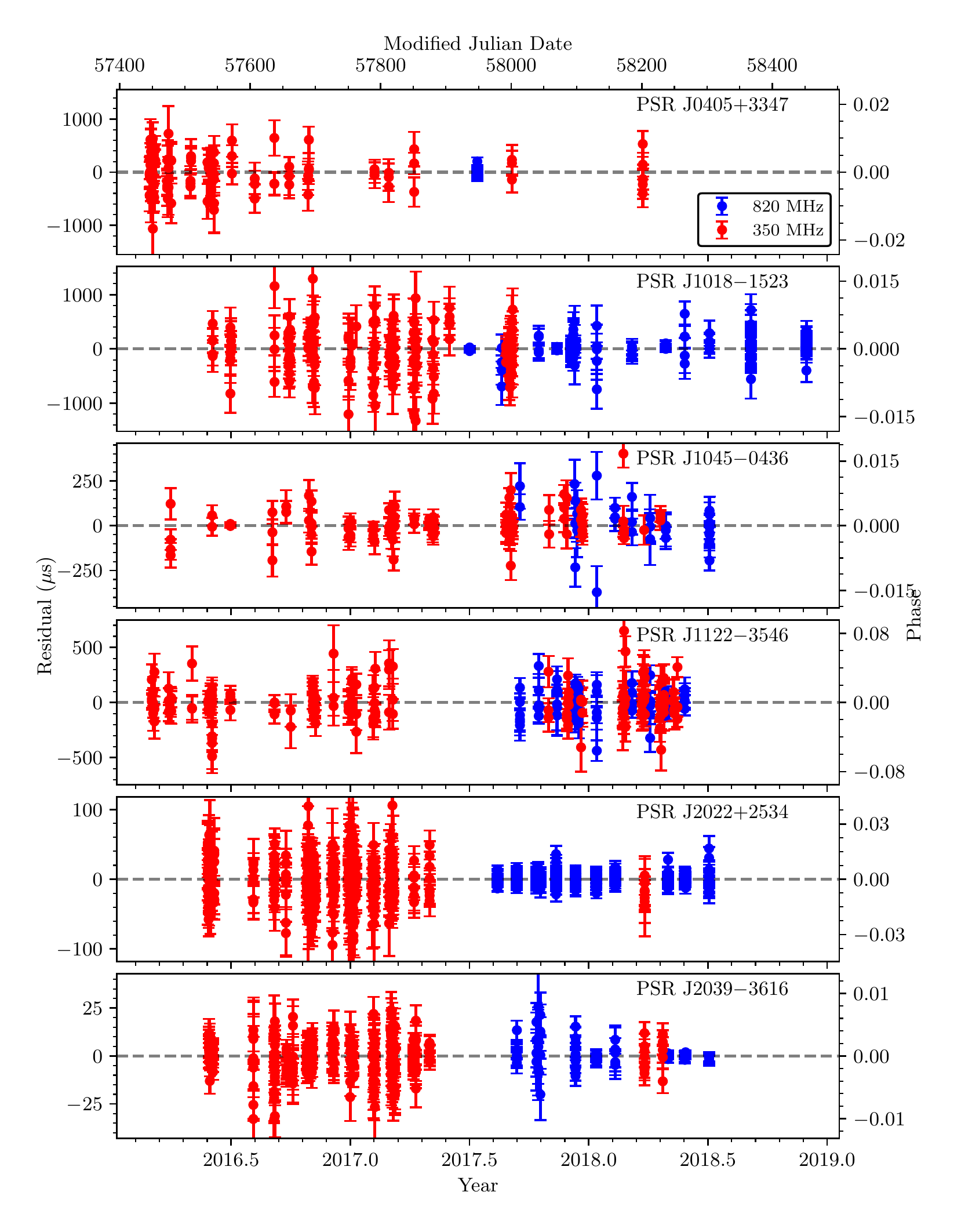}
\centering
\caption{Timing residuals for PSRs~\psra, \psrc, \psrd, \psre, \psrk, and \psrl, from observations at 350\,MHz (red) and 820\,MHz (blue) respectively. Error bars represent 1-$\sigma$ uncertainties on individual TOA measurements. Note: timing residuals for PSR \psrb are not plotted here since those data come from an earlier study (GBT project code 15A$-$376; PI: L.\ Levin) and span a very different period of time.}
\label{fig:resids}
\end{figure*}

Final sets of timing residuals (differences between measured/expected times of arrival) from this refinement process are plotted in Figure \ref{fig:resids}. Results from fitting for spin and astrometric parameters are listed in Tables \ref{tab:spin} and \ref{tab:astrometric} respectively, and corresponding derived parameters can be found in Table \ref{tab:derived}. In some cases, the reduced $\chi^2$ values were relatively far from 1, so TOA uncertainties have been scaled by multiplicative error factors (EFACs; see Table \ref{tab:spin}) to force $\chi_{\rm red}^2=1$. This scaling also impacts uncertainties on parameters measured via pulsar timing.


For three MSPs (PSRs~\psrb, \psrk, and \psrl), timing precision was sufficient to measure their proper motions in ecliptic longitude and latitude,  $\mu_\lambda$ and $\mu_\beta$. Proper motion is detectable with pulsar timing and manifests itself as a growing sinusoid in the pulsar timing residuals. Typically this signature is only detectable after timing pulsars over longer timespans ($\gtrsim$\,3 years) or for young pulsars that have substantial kick velocities, but proper motion can also be detected over 1$-$2 year timespans for nearby MSPs with high-precision TOAs, as is the case here.

Total proper motions and DM distances ($D_{\rm DM}$)  for these MSPs were used to compute transverse velocities ($v_{\rm t}$; see Table \ref{tab:pms}). Transverse motion translates to an apparent spindown due to a pulsar's motion relative to the solar system barycenter; this is called the Shklovskii effect \citep{shklovskii+70}, \ps, and is typically only significant for nearby MSPs whose $\dot{P}$  values already tend to be small. A pulsar's acceleration in the Galactic potential can also contribute to the measured spindown. However, this factor, \pg, is usually only significant for relatively distant MSPs. We follow the same procedure as described in \cite{gfg+21} to calculate \pg, which includes an approximation for the vertical component of Galactic acceleration \citep{hf+04} and the latest values for the distance between the sun and Galactic center and the circular velocity of the Sun \citep[$R_0=8.275\pm0.034$\,kpc and $\Phi_0=240.5\pm4.1$\,km\,s$^{-1}$, respectively;][]{GC+21,rb+20}. In Table \ref{tab:pms}, we calculate $\mu_\lambda$, $\mu_\beta$, and $v_t$ for PSRs~\psrb, \psrk, and \psrl, then use NE2001 \citep{cl+02} and YMW16 \citep{ymw+17} distance models\footnote{Also see \url{https://pulsar.cgca-hub.org/compute}.} to determine intrinsic spindown values, \pint, by subtracting \ps and \pg components from the measured $\dot{P}$. Finally, we compute resulting surface magnetic field ($B_{\rm surf}$), characteristic age ($\tau_{\rm c}$), and spindown luminosity ($\dot{E}$) using each pulsar's measured spin period, $P$, and \pint.

For most of the binary systems presented here, we used the ELL1 timing model \citep{lcw+01}, which parameterizes orbital parameters in terms of the epoch of the ascending node ($T_{\rm asc}$) and first and second Laplace-Lagrange parameters ($\epsilon_1$ and $\epsilon_2$). This is a convenient prescription for low-eccentricity systems with short-period orbits.  For \psrc, which we suspect is a new DNS system, we employed the DD model \citep{dd+86} and fit for one relativistic parameter (advance of periastron, $\dot{\omega}$), in addition to the usual five Keplerian parameters used to describe binary orbits. Results of binary parameter fits can be found in Tables \ref{tab:dns} and \ref{tab:binary}.

In several cases, TOAs from discovery scans were included to improve spindown ($\dot{\nu}$) measurements (e.g. see PSRs~\psrg, \psrh, \psri, \psrj). A Taylor expansion was used to express each pulsar's expected phase as a function of time, given initial measurements for spin frequency and frequency derivative and their uncertainties from our timing campaign. In each case where discovery TOAs were included, we ensured that pulse phase uncertainties accumulated over $3-6$ month gaps amounted to $\ll1$ rotation. Additional analysis and interpretation of timing models and parameters measured for individual sources can be found in Section \ref{sec:disc}.


\begin{deluxetable*}{llrllrcc}
  \tabletypesize{\footnotesize}
  \tablewidth{\textwidth}
  \tablecaption{Derived Common Properties of GBNCC Pulsars}
  \tablecolumns{8}
  \tablehead{
    \colhead{PSR} & \colhead{$P$} & \colhead{$\dot{P}$} &
    \colhead{$\tau_{\rm c}$} & \colhead{$B_{\rm surf}$} &
    \colhead{$\dot{E}$} & \colhead{D$^{\rm NE2001}_{\rm DM}$} &
    \colhead{D$^{\rm YMW16}_{\rm DM}$} \\
    \colhead{} & \colhead{(s)} & \colhead{(s\,s$^{-1}$)} &
    \colhead{(yr)} &  \colhead{(Gauss)} & 
    \colhead{(erg\,s$^{-1}$)} & \colhead{(kpc)} & \colhead{(kpc)}}
    \startdata
J0405+3347 & 0.0639539495051(1) & 1.7(1)$\times10^{-19}$ & 6.1$\times10^{9}$ & 3.3$\times10^{9}$ & 2.5$\times10^{31}$ & 1.9 & 1.7 \\
J0742+4110 & 0.0031391369352908(3) & 6.692(5)$\times10^{-21}$ & 7.4$\times10^{9}$ & 1.5$\times10^{8}$ & 8.5$\times10^{33}$ & 0.7 & 0.5 \\
J1018$-$1523 & 0.0831525352278(3) & 1.09(6)$\times10^{-19}$ & 1.2$\times10^{10}$ & 3.0$\times10^{9}$ & 7.5$\times10^{30}$ & 0.8 & 1.1 \\
J1045$-$0436 & 0.02404751703698(1) & 7.8(1)$\times10^{-20}$ & 4.9$\times10^{9}$ & 1.4$\times10^{9}$ & 2.2$\times10^{32}$ & 0.3 & 0.3 \\
J1122$-$3546 & 0.007838069357439(3) & 1.53(7)$\times10^{-20}$ & 8.1$\times10^{9}$ & 3.5$\times10^{8}$ & 1.3$\times10^{33}$ & 1.5 & 0.7 \\
J1221$-$0633 & 0.0019345396958571(2) & 5.25(8)$\times10^{-21}$ & 5.8$\times10^{9}$ & 1.0$\times10^{8}$ & 2.9$\times10^{34}$ & 0.8 & 1.2 \\
J1317$-$0157 & 0.002908244599817(2) & 5.4(5)$\times10^{-21}$ & 8.6$\times10^{9}$ & 1.3$\times10^{8}$ & 8.6$\times10^{33}$ & 2.8 & 25.0\phn \\
J1742$-$0203 & 0.131609685521(9) & 1.5(1)$\times10^{-16}$ & 1.4$\times10^{7}$ & 1.4$\times10^{11}$ & 2.6$\times10^{33}$ & 2.8 & 3.7 \\
J2017$-$2737 & 0.22452428375(9) & 6.1(1)$\times10^{-15}$ & 5.8$\times10^{5}$ & 1.2$\times10^{12}$ & 2.1$\times10^{34}$ & 1.0 & 1.6 \\
J2018$-$0414 & 0.0406122096643(1) & 6(4)$\times10^{-20}$ & 1.1$\times10^{10}$ & 1.5$\times10^{9}$ & 3.3$\times10^{31}$ & 1.5 & 1.8 \\
J2022+2534 & 0.0026459357677905(3) & 6.16(1)$\times10^{-21}$ & 6.8$\times10^{9}$ & 1.3$\times10^{8}$ & 1.3$\times10^{34}$ & 3.3 & 4.0 \\
J2039$-$3616 & 0.0032750415513069(2) & 8.427(9)$\times10^{-21}$ & 6.2$\times10^{9}$ & 1.7$\times10^{8}$ & 9.5$\times10^{33}$ & 0.9 & 1.7 \\
  \enddata
  \tablecomments{D$_{\rm DM}$ is calculated using the NE2001
  \citep{cl+02} or YMW16 \citep{ymw+17} Galactic free electron density
  models, as indicated.  A fractional error of 50\% is not
  uncommon.  Derived parameters here have not been corrected for apparent acceleration caused by kinematic effects.  $\dot{E}$ and $B_{\rm surf}$ are calculated assuming a moment of inertia $I = 10^{45}$\,g\,cm$^2$; additionally, $B_{\rm surf}$ assumes a neutron star radius $R=10$\,km and $\alpha=90^{\circ}$ (angle between spin/magnetic axes). Calculating $\tau_{\rm c}$ relies on the assumption that spin-down is fully due to magnetic dipole radiation (braking index, $n=3$) and that the initial spin period is negligible. Values in parentheses are the $1$-$\sigma$ uncertainty in the last digit, calculated by propagating uncertainties in measured parameters reported by
  \texttt{TEMPO}.}
  \end{deluxetable*}
\label{tab:derived}

\begin{deluxetable*}{lllcccccccc}
  \centering
  \tabletypesize{\footnotesize}
  \tablewidth{0pt}
  \tablecolumns{11}
  \tablecaption{Proper Motions and Kinematic Corrections for Three GBNCC Pulsars}
  \tablehead{
    \colhead{PSR}                                    &
    \colhead{$\mu_\lambda$}     &
    \colhead{$\mu_\beta$}       &
    \colhead{$D_{\rm DM}$} &
    \colhead{$v_{\rm t}$}            &
    \colhead{\pg}    &
    \colhead{\ps}    &
    \colhead{\pint}   &
    \colhead{$B_{\rm surf}$} &
    \colhead{$\tau_{\rm c}$} &
    \colhead{$\dot{E}$} \\
    \colhead{}                      &
    \colhead{($\mathrm{mas}\, \yr^{-1}$)} &
    \colhead{($\mathrm{mas}\, \yr^{-1}$)} &
    \colhead{(\kpc)} &
    \colhead{($\km\, \s^{-1}$)}      &
    \colhead{($10^{-21}$)}           &
    \colhead{($10^{-21}$)}           &
    \colhead{($10^{-21}$)}           &
    \colhead{($10^8\; \gauss$)} &
    \colhead{(Gyr)} &
    \colhead{($10^{33}\; \erg\, \s^{-1}$)}
  }
  \startdata
J0742+4110 & $-$12(2) & $-$9(5) & 0.7(2) & 5(2)$\times10^{1}$ & $-$0.01 & 1.27 & 5.43 & 1.3 & 9.2 & 6.9 \\
 & & & 0.5(2) & 4(1)$\times 10^{1}$ & $-$0.02 & 0.93 & 5.79 & 1.4 & 8.6 & 7.4 \\
J2022+2534 & \phn$-$4.0(7) & $-$8(1) & 3(1) & 1.4(5)$\times10^{2}$ & $-$0.83 & 1.77 & 5.22 & 1.2 & 8.0 & 11.1\phn \\
 & & & 4(1) & 1.7(6)$\times 10^{2}$ & $-$1.03 & 2.14 & 5.05 & 1.2 & 8.3 & 10.8\phn \\
J2039$-$3616 & $-$13.5(4) & \phnd2(1) & 0.9(3) & 6(2)$\times10^{1}$ & $-$0.11 & 1.35 & 7.19 & 1.6 & 7.2 & 8.1 \\
 & & & 1.7(5) & 1.1(3)$\times10^{2}$ & \phnd0.00 & 2.51 & 5.92 & 1.4 & 8.8 & 6.7 \\
  \enddata
  \tablecomments{For each pulsar in this study with measurable proper motion, we list measurements in ecliptic longitude and latitude ($\mu_\lambda$ and $\mu_\beta$). Distances estimated using pulsars' DMs and Galactic electron density models ({\it top:} NE2001, {\it bottom:} YMW16) are quoted with $\approx30\%$ uncertainty. Calculating transverse velocity ($v_t$) based on these quantities allow us to calculate, in turn, secular acceleration (Shklovskii effect; \ps) and that due to pulsars' motion in the Galactic potential (\pg); removing these factors from $\dot{P}$ gives the intrinsic value, \pint, which is used to calculate derived quantities, surface magnetic field strength ($B_{\rm surf}$), characteristic age ($\tau_{\rm c}$), and spin-down luminosity ($\dot{E}$).}
\end{deluxetable*}
\label{tab:pms}

\renewcommand{\thefigure}{\arabic{figure} (cont.)}
\addtocounter{figure}{-1}

\begin{figure*}[!htb]
\centering
\includegraphics[width=0.9\textwidth]{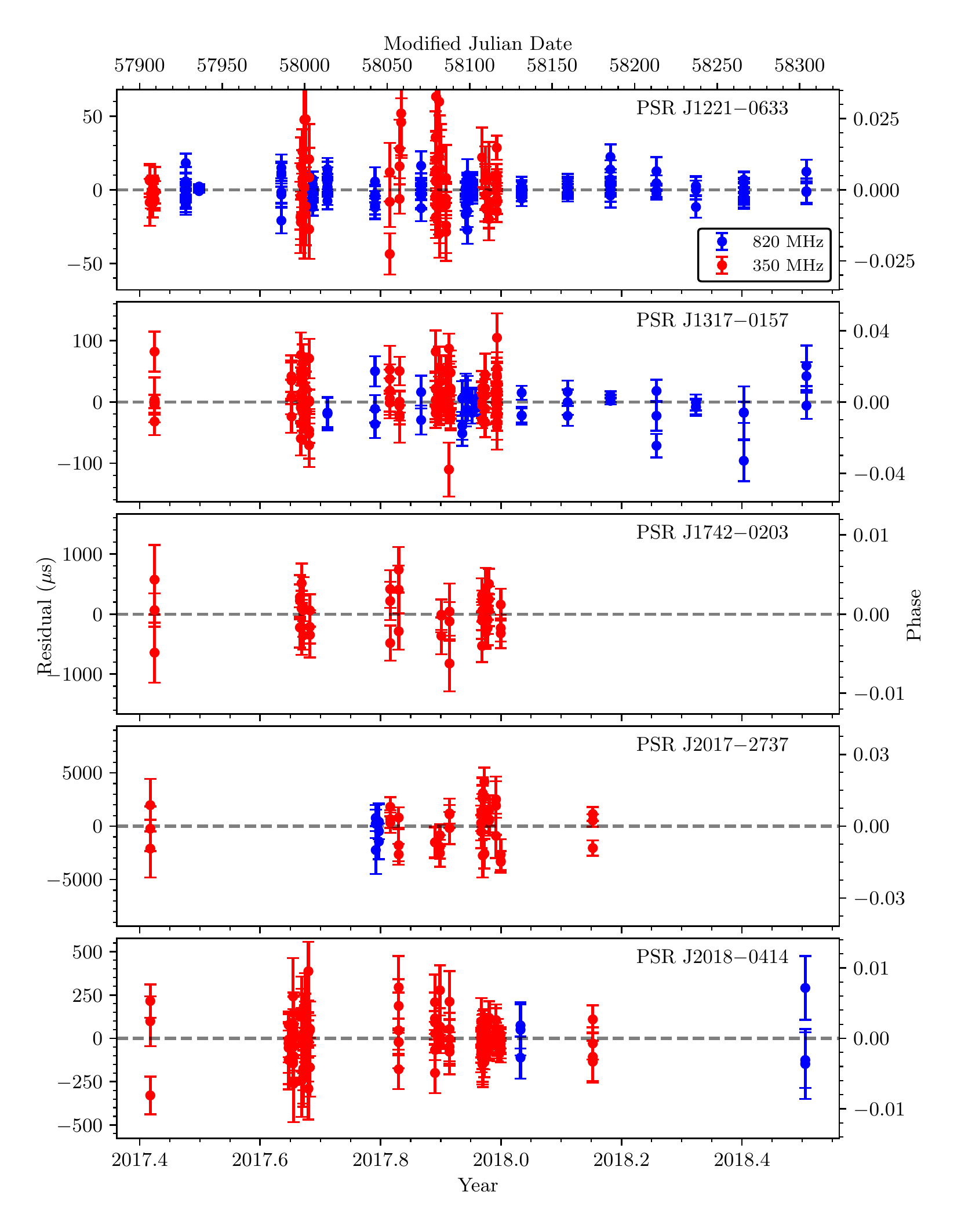}
\centering
\caption{Timing residuals for PSRs~\psrf, \psrg, \psrh, \psri, and \psrj, from observations at 350\,MHz (red) and 820\,MHz (blue) respectively. Error bars represent 1-$\sigma$ uncertainties on individual TOA measurements.}
\end{figure*}

\renewcommand{\thefigure}{\arabic{figure}}

\subsection{Fermi-LAT counterparts}\label{sec:fermi}

The last decade has seen the discovery of a profusion of $\gamma$-ray pulsars\footnote{See \url{http://tinyurl.com/fermipulsars} for an overview.} thanks to the Large Area Telescope \citep[LAT;][]{aaa+09} on the \textit{Fermi Gamma-ray Space Telescope}. LAT has been continuously imaging the sky in the energy range from $\sim$20 MeV to 1 TeV since 2008. 

Several pulsars in that rich dataset have been identified through deep radio searches targeting \textit{Fermi}-LAT unidentified point sources \citep[e.g.,][]{rrc+11,kcj+12,rap+12,ckr+15,cck+16,pbh+17,drc+21}. However, high levels of background contamination -- particularly in the Galactic plane, where most pulsars reside -- may cause $\gamma$-ray pulsars to be confused and undetectable as point sources. An alternative approach that has proven fruitful is by selecting $\gamma$-ray photons coming from the position of known radio pulsars and phase-folding the data using coherent timing solutions derived from the radio data \citep[e.g.,][]{aaa+10,aaa+13,sbc+19}. In this work, we have searched for high-energy counterparts to our pulsars through both identification methods. 

We inspected the Fermi Large Array Telescope (LAT) 12-year $\gamma$-ray source catalog \citep[4FGL-DR3;][]{aab+22} to identify objects spatially coincident with the timing positions of the pulsars we derived from the GBT data. Three of the 12 pulsars have positions coincident with \textit{Fermi} point sources. The timing positions of PSRs~\psrf and \psrl are within the 68\% confidence region of the sources 4FGL~J1221.4$-$0634 (detection significance of 23.4\,$\sigma$) and 4FGL~J2039.4$-$3616 (15.2\,$\sigma$), respectively. Both 4FGL sources have pulsar-like power-law spectra with subexponential cutoffs. The third potential association is PSR~\psrg co-located within the 95\% confidence region of 4FGL J1317.5$-$0153 (9.4-$\sigma$ detection). This source has a log-normal spectrum, which is not as common as the subexponential cutoff power law among known $\gamma$-ray pulsars. The likelihood of the 4FGL point sources being counterparts to the pulsars in terms of pulsar energetic and \textit{Fermi}-LAT sensitivity is examined further below. But first, we describe the method we used to search for high-energy pulsed emission through the phase-folding of the \textit{Fermi}-LAT photons.  

For all pulsars, we retrieved LAT photons\footnote{Data were downloaded from the LAT Data server available here: \url{https://fermi.gsfc.nasa.gov/cgi-bin/ssc/LAT/LATDataQuery.cgi}} within 
3$^\circ$ of the timing positions collected from 2008 August 5 ($\approx$ the start of the mission) to 2022 August 5. We selected photons having an energy $E_\gamma$ in the range $0.1<E_\gamma<500$\,GeV and applied the standard events screening recommended by the \textit{Fermi}-LAT team\footnote{Pass 8 data analysis \citep{bbd+18}, see also the \textit{Fermi} Science Support Center; \url{https://fermi.gsfc.nasa.gov/ssc/data/analysis/documentation/Pass8_usage.html}}. Good time intervals (where the telescope observed nominally) were selected using \texttt{gtmktime} and photon arrival times were corrected to the Earth's geocenter with the \texttt{gtbary} tool.

Using PINT's \texttt{fermiphase}\footnote{\url{https://github.com/nanograv/PINT}} tool, we assigned to each photon a probability of being emitted by the pulsar (i.e., weighted) following the method from \citet{b19}. Weight computations are based on the photon energy and angular separation from the target position for an assumed spectral distribution. The only free parameter in the weight model is the 
$\mu_E=\log_{10}\left(E_{\rm ref}/1\rm{MeV}\right)$,
where $E_{\rm ref}$ is the reference energy at which the distribution of photon weights peaks (see Eq. 11 of \citealt{b19}). The bulk of the known $\gamma$-ray pulsars have $\mu_E\approx$\,3.6 (equivalently, $E_{\rm ref}\approx4$\,GeV), but hard-spectrum sources in highly confused regions may favor $\mu_E>4$. 

Considering that the pulsars in this work are located at various Galactic latitudes and therefore are subject to different types of background contamination, we phase-folded the LAT dataset with four trial $\mu_E$ in the weight calculation, with $\mu_E \in \{2.8, 3.2, 3.6, 4.0\}$. The weighted H-test statistic \citep{jb10,b19} was calculated to assess the significance of the pulsations. An additional filtering of low-weight photons was then applied in order to identify the minimum photon weight $w_{\rm min}$ that maximizes the H-test (to value $H_{\rm max}$) for each trial $\mu_E$. To avoid potential sensitivity losses due to long-term timing effects (e.g., proper motion) that are not modeled in the timing solutions, we repeated the same procedure but this time selecting only events within the validity range of the radio ephemerides.

Following the analysis of \citet{sbc+19}, who used a similar approach to fold over a thousand pulsars and examine their H-test distribution to identify the ideal selection criteria to reject false positives, we dismissed candidates having $H_{\rm max}<25$ (equivalent to a $\approx$4-$\sigma$ detection) across all trial combinations ($w$, $\mu_E$). Further optimization was performed for statistically significant detections by performing a finer search in trial $\mu_E$. 

Among the 12 pulsars presented in this work, pulsations were detected in two pulsars, PSRs~\psrf and \psrl, which are two of the three pulsars that are co-located with 4FGL sources. These, along with the non-detection of pulsations in PSR~\psrg co-located with 4FGL J1317.5$-$0153, are discussed below. Apart from PSRs~\psrf and \psrl, no significant pulsations were detected in the four other pulsars (PSRs~\psrb, \psrd, \psre, \psri) that have ``heuristic'' energy fluxes, $G_h\,=\,\sqrt{\dot{E}}/4\pi d^2$, above the typical LAT detection threshold of 10$^{15}$\,(erg\,s$^{-1}$)$^{1/2}$\,kpc$^{-2}$ \citep{aaa+13,sbc+19}. This suggests that their DM-inferred distances could be underestimated, and/or that any beam of high-energy photons emitted by these pulsars do not intercept our line of sight. The latter is further supported by the large pulse width of these pulsars in the radio (all have duty cycles, $\delta>0.3$, see Table \ref{tab:flux}), which is generally indicative a low magnetic inclination and empirically associated with non-detection of $\gamma$-ray pulsations \citep[see e.g.,][]{rwjk17,sbc+19,jskk20,sjk+21}.    

\subsubsection{PSR~\psrf}\label{sec:fermi-J1221}
PSR~\psrf is spatially coincident with the bright \textit{Fermi} source 4FGL~J1221.4$-$0634. Here we consider the spindown power of the pulsar to determine if the properties of the coincident 4FGL object are consistent with being the counterpart of PSR~\psrf. The pulsar has a spindown power of $\dot{E}=2.9\times10^{34}$\,erg\,s$^{-1}$, and assuming an average DM distance of 1\,kpc, the corresponding heuristic flux $G_h\sim10^{16}$\,(erg\,s$^{-1}$)$^{1/2}$\,kpc$^{-2}$ is well above the LAT threshold. The 4FGL-DR3 reports an integrated energy flux in the 0.1--100\,GeV band, $G_{100}$, of $5.8\times10^{-12}$\,erg\,s$^{-1}$\,cm$^{-2}$ for 4FGL~J1221.4$-$0634. Assuming a $\gamma$-ray beaming fraction $f_\Omega=1$ (appropriate for outer-magnetosphere emission sweeping a full 4$\pi$ steradians), the luminosity $L_\gamma$ of the \textit{Fermi} source ranges between $0.4-1\times10^{33}$\,erg\,s$^{-1}$, depending on the adopted DM distance. Comparing the power radiated in the 0.1--100\,GeV band to the spindown power, the $\gamma$-ray conversion efficiency $\eta=L_\gamma/\dot{E}$ of 4FGL~J1221.4$-$0634 is between 15 and 35\%. These results are all consistent with 4FGL~J1221.4$-$0634 being the counterpart of PSR~\psrf. 

\begin{figure}
\centering
\includegraphics[scale=0.65]{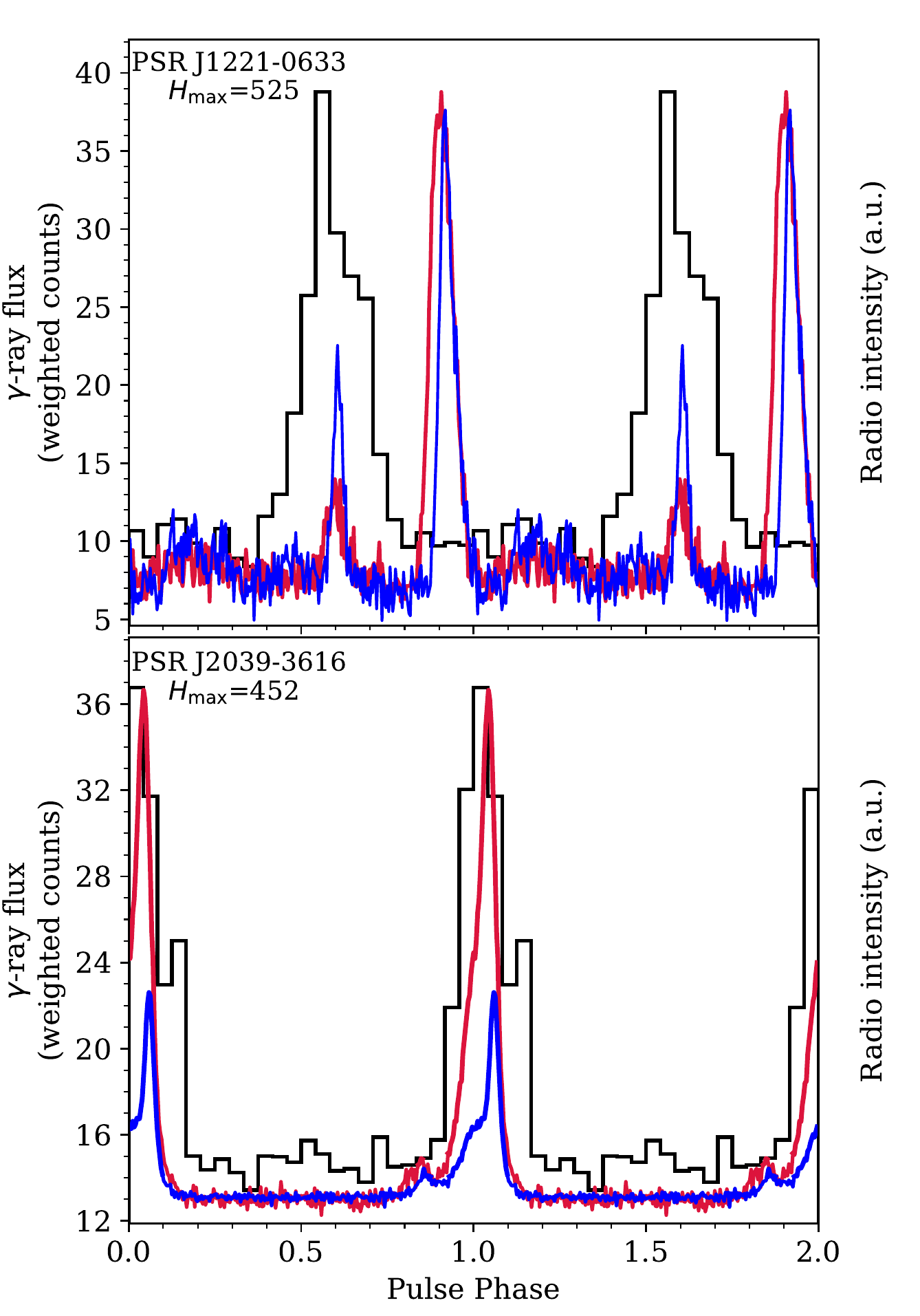}
\caption{$\gamma$-ray (black) and radio pulse profiles at 350\,MHz (red) and 820\,MHz (blue) for PSRs~\psrf (top) and \psrl (bottom), phase-aligned with their respective timing ephemerides (Section~\ref{sec:fits}). The $\gamma$-ray profiles were generated using the photon filtering and energy scaling that maximized the $H$-test value (see $H_{\rm max}$ in top-left of each panel), as described in Section~\ref{sec:fermi}. All radio profiles are shown with 256 phase bins per cycle, while the $\gamma$-ray profiles have 24 phase bins per cycle.}
\label{fig:gray_profs}
\end{figure}

Phase-folding the \textit{Fermi} photons collected in the direction of PSR~\psrf resulted in strong pulsations ($H_{\rm{max}}\,=\,525$). The weight model that optimized the significance of the pulse profile had an energy scale $\mu_E\,=\,3.7$ ($E_{\rm ref}\approx5$\,GeV) and photons with $w>1$\%. 
Figure~\ref{fig:gray_profs} shows the binned $\gamma$-ray pulse profile overlaid with the profiles of the pulsar at 350 and 820\,MHz after barycentering the arrival times in both bands and correcting  for time delays due to ISM propagation effects. The final timing ephemeris (Table~\ref{tab:spin}) was used to calibrate the absolute phase alignment to the same reference time and frequency. In the radio band, PSR~\psrf displays two distinct peaks separated by $\sim110^\circ$ (or equivalently 0.31 in rotational phase), whereas only one broad component (pulse duty cycle $\delta$ of 0.18/0.35 at 50\%/10\% of the peak maximum intensity) is seen in the $\gamma$-ray profile at the same rotational phase as the leading (and weaker) radio peak\footnote{We note that we used the zero-phase reference epoch in the timing solution as the fiducial phase. If instead we adopt the same approach as \cite{aaa+13} and set the phase of the peak radio intensity as the fiducial phase, then the $\gamma$-ray and weaker radio peaks of PSR~\psrf seen in Figure~\ref{fig:gray_profs} is trailing the main radio peak by a phase of 0.69.}. This is consistent with the $\gamma$-rays and fainter radio beam being produced at a similar altitude \citep[see e.g.,][]{jvh+14}. 
These results further support the association of 4FGL~J1221.4$-$0634 with PSR~\psrf. \\

\subsubsection{PSR~\psrg}\label{sec:fermi-J1317}
As previously mentioned, the point source 4FGL J1317.5$-$0153 is coincident with PSR~\psrg yet no $\gamma$-ray pulsations were detected in the folded \textit{Fermi} data. Despite having a spindown luminosity ($\dot{E}\sim9\times10^{33}$\,erg\,s$^{-1}$) above the empirical $\gamma$-ray emission deathline for MSPs ($\dot{E}_{\rm death}=8\times10^{32}$\,erg\,s$^{-1}$; \citealt{gsl+16}), the large (and highly uncertain) distance predicted by the YMW16 model based on the DM along the pulsar line of sight (D$^{\rm YMW16}_{\rm DM}\,>$\,25\,kpc,  i.e., exceeding the maximum Galactic contribution to the DM in that direction) translates into an energy flux $G_h$ that is two orders of magnitude below the expected LAT sensitivity. To meet the LAT detectability threshold, the distance of the pulsar should be within $\sim$2.75\,kpc, which is consistent with the distance predicted by the NE2001 model (D$^{\rm NE2001}_{\rm DM}$\,=\,2.8\,kpc). The \textit{Fermi} point source has a 0.1--100\,GeV flux of $G_{100}=1.5\times10^{-12}$\,erg\,s$^{-1}$\,cm$^{-2}$. Assuming a beaming fraction $f_\Omega=1$, the $\gamma$-ray luminosity of 4FGL J1317.5$-$0153 at the NE2001 distance is $L_\gamma = 1.4\times10^{33}$\,erg\,s$^{-1}$. This translates into a $\gamma$-ray efficiency $\eta=16$\%, a typical value among known $\gamma$-ray MSPs \citep[see e.g.,][]{aaa+13,sbc+19}. Apart from the overestimated/unconstrained YMW16 distance, the association of 4FGL J1317.5$-$0153 with PSR~\psrg is reasonable in terms of the expected \textit{Fermi}-LAT sensitivity, the pulsar spindown luminosity and the low background level in that sky region. An unfavorable viewing geometry and/or magnetic alignment could explain the non-detection of $\gamma$-ray pulsations in this pulsar. Careful modeling of the radio profile evolution and polarization properties could help determine the validity of the magnetosphere-geometry argument for the non-detection at high energies \citep[e.g.,][]{rwjk17}. Such analysis is however beyond the scope of this work. It should also be noted that the 4FGL-DR3 catalog reports a 50\% probability that this \textit{Fermi} source is associated with the active galactic nucleus CRATES~J1317$-$0159 \citep{hrt+07}. At this point, we cannot conclusively associate 4FGL J1317.5$-$0153 with PSR~\psrg.

\subsubsection{PSR~\psrl}\label{sec:fermi-J2039}
After correcting for the apparent accelerations that arise from kinematic effects ( Section~\ref{sec:ta}), the intrinsic spindown power $\dot{E}$ and heuristic energy flux $G_h$ we estimated for PSR~\psrl are $\dot{E} = 6.7\times 10^{33}$\,erg\,s$^{-1}$ and $G_h=2\times 10^{15}$\,(erg\,s$^{-1}$)$^{1/2}$\,kpc$^{-2}$ when adopting the distance predicted by the YMW16 model (D$^{\rm YMW}_{\rm DM}$=1.7\,kpc), and $\dot{E} = 8.1\times 10^{33}$\,erg\,s$^{-1}$ and $G_h=9\times 10^{15}$\,(erg\,s$^{-1}$)$^{1/2}$\,kpc$^{-2}$ for the NE2001 distance (D$^{\rm NE2001}_{\rm DM}$=0.9\,kpc). Both distance predictions are small enough to produce fluxes above the LAT sensitivity to point sources -- in fact, LAT detectability ($G_h > 10^{15}$\,(erg\,s$^{-1}$)$^{1/2}$\,kpc$^{-2}$) is ensured for a pulsar distance $<$\,2.6\,kpc.  

The \textit{Fermi} source coincident with \psrl, 4FGL~J2039.4$-$3616, has a energy flux of $G_{100}=3.8\times10^{-12}$\,erg\,s$^{-1}$\,cm$^{-2}$. When the distance predicted by NE2001 is adopted, the corresponding luminosity and efficiencies are $L_\gamma=0.4\times10^{33}$\,erg\,s$^{-1}$ and $\eta$ = 5\%. If instead the larger distance predicted YMW16 is used, we obtain  $L_\gamma=1.3\times10^{33}$\,erg\,s$^{-1}$ and $\eta$ = 20\%. 

We also detected bright $\gamma$-ray pulsed emission from PSR~\psrl when we phase folded the \textit{Fermi} photons. Filtering out photons with $w<0.5$\% and setting $\mu_E=3.75$ yielded the strongest pulsations at a significance of $H_{\rm{max}}=452$. 
The $\gamma$-ray pulse profile of PSR~\psrl is relatively narrow ($\delta$ of 0.12/0.28 at 50\%/10\% of the peak intensity) and single-peaked, and is aligned with the rotational phase of the main peak of the (complex) profile in both radio bands (see Figure~\ref{fig:gray_profs}). Phase alignment of the radio and $\gamma$-ray beams suggests the co-location of emission regions across waveband \citep[][]{aaa+10b}, possibly high in altitude and caustic in origin \citep[][]{vjh12,jvh+14}. PSR~\psrl is an interesting target for testing emission geometry models -- radio polarization information would be most helpful, however only total-intensity data were recorded on this pulsar for this project.

In light of the properties discussed above and the firm detection of pulsed GeV emission, we identify PSR~\psrl as the source powering 4FGL~J2039.4$-$3616.  


\section{Discussion}\label{sec:disc}
\subsection{Isolated Pulsars}
Five pulsars (PSRs~\psra, \psre, \psrh, \psri, and \psrj) were included in this study based on their relatively high spin frequencies, possibly indicative of spin-up due to a previous recycling period, however two show no signs of recycling (PSRs~\psrh and \psri appear to be young, canonical pulsars) and the other three do, but are no longer bound to their binary companions.

\subsubsection{Non-Recycled Pulsars}
After its discovery, PSR~\psrh was considered a candidate binary pulsar due to its intermediate, 132\,ms spin period. However, extended timing showed no evidence of the pulsar being in a binary system and given its measured period derivative ($1.5\times10^{-16}$\,s/s), it is also unlikely to be recycled. Although only a single attempt was made, PSR~\psrh was not detected at 820\,MHz in a 5\,min scan. Using the non-detection, we place an apparent upper limit on its flux density at 820\,MHz, $S_{820}<0.2$\,mJy, using the radiometer equation (assuming a duty cycle, $\delta=0.06$ and signal-to-noise detection threshold of ${\rm S/N}>6$), which also constrains its spectral index, $\alpha<-1.5$. Since PSR~\psrh was never detected at 820\,MHz, the DM value listed in Table \ref{tab:astrometric} maximizes S/N in a high-S/N 350 MHz observation; we were not able to reliably fit for DM in {\tt TEMPO} with single-frequency data, so we do not include a corresponding uncertainty in the table.

Similar to PSR~\psrh, PSR~\psri was considered as a candidate recycled pulsar due to its intermediate spin period, 225\,ms. However, concerted timing efforts showed that its $\dot{P}$, and therefore, derived $B_{\rm surf}$ values were too large to indicate a previous period of recycling.

\subsubsection{Disrupted Recycled Pulsars}
PSR~\psra is a solitary pulsar with a relatively short, 64\,ms spin period and weak magnetic field ($B_{\rm surf}=3.3\times10^9$\,G). Using the definition for a disrtupted recycled pulsar (DRP) posed by \cite{blr+10} -- an isolated pulsar in the Galactic disc, with $B<3\times10^{10}$\,G and $P>20$\,ms -- PSR~\psra is a new DRP. Its spin parameters indicate that it is partially recycled, but the recycling process was likely cut short when its companion went supernova, disrupting the system and producing an isolated, young pulsar and a partially recycled isolated pulsar \citep[e.g., see discussion of J1821+0155 in][]{rsm+13}. Although its period derivative is not as well constrained, PSR~\psrj too has a short, 41\,ms spin period and likely evolved via the same mechanism as PSR~\psra. 

\subsubsection{PSR~\psre}
PSR~\psre is an isolated MSP with a 7.8\,ms spin period. It is widely accepted that MSPs have such short spin periods as a result of an extended recycling process, where material from a binary companion transfers angular momentum, ``spinning up" the pulsar \citep{alpar+82}. Consistent with this evolutionary theory, we find the majority of MSPs reside in near-circular orbit binary systems with low mass white dwarf (WD) companions.\footnote{See e.g., \url{http://www.atnf.csiro.au/research/pulsar/psrcat} \citep{mht+05}.} Isolated MSPs like PSR~\psre are comparatively uncommon, comprising roughly 25\% of the MSP population in the Galactic field, but their evolutionary history remains an open question.

\subsection{MSPs Suitable for PTAs}
The primary science goal for the GBNCC pulsar survey is discovering new MSPs, particularly those suitable for high-precision timing in the effort to use PTAs to detect low-frequency GWs. Generally, bright MSPs with short spin periods and sharp features in their profile are best, ideally producing TOAs with timing residuals that have root mean square (RMS) $<\,1$\,\us (at observing frequencies $\gtrsim\,1$\,GHz). Both PSRs~\psrk and \psrl satisfy these basic criteria, and therefore, have been added to the NANOGrav PTA for use in low frequency GW detection and characterization.

PSR~\psrk is a 2.6\,ms pulsar in a short, 1.3\,day circular orbit around a low mass ($m_{\rm c,min}=0.07$\,\Msun) companion. Its eccentricity is close to zero, which is typical for MSPs, whose long recycling periods tend to circularize their orbits \citep{pk+94}. Although this pulsar appears to be bright across the frequency spectrum between 300\,MHz and 2.5\,GHz,\footnote{Higher frequency testing done at Arecibo Observatory by Andrew Seymour, private communication.} its profile is broad (duty cycle, $\delta>0.5$) and lacks narrow features at 350\,MHz, which makes timing imprecise at low observing frequencies. Figure \ref{fig:resids} shows a comparison between TOA precision at 350\,MHz versus 820\,MHz. At 820\,MHz and higher observing frequencies, PSR~\psrk remains bright and although its average profile envelope is still broad, sharp features provide consistent anchors for profile template matching, making it a promising candidate for PTA science. 

PSR~\psrk has been added to the NANOGrav PTA -- initially monitored at the Arecibo Observatory, but now regularly at the Green Bank Observatory and with the Canadian Hydrogen Intensity Mapping Experiment \citep[CHIME; ][]{chime+21} telescope -- and it will continued to be monitored closely in the coming years to ensure long-term timing stability necessary for MSPs used to detect GWs.

PSR~\psrl is a 3.3\,ms pulsar in a 5.8\,day, nearly circular orbit about a (likely) low mass WD companion. Like PSR~\psrk, PSR~\psrl remains bright up to observing frequencies of 2.5\,GHz\footnote{Higher frequency testing done at Green Bank Observatory by the NANOGrav Collaboration, private communication.} and its sharp profile makes it a good PTA candidate; observing at 820\,MHz and above, individual TOA uncertainties are typically $<1$\,\us, so PSR~\psrl has been included in NANOGrav's PTA. PSR~\psrl is near Green Bank Observatory's low declination limit (GBO can observe sources with ${\rm Dec.}\gtrsim-45^{\circ}$), so this pulsar may also prove useful for PTA experiments with better access to southern hemisphere sources, like the PPTA \citep{rsc+21}, InPTA \citep{jab+18}, and MeerTime (Miles et al. in press).

For PSR~\psrl, in addition to proper motion (see Table \ref{tab:astrometric}), we find a significant measurement of parallax $5.5\pm1.2$\,mas, implying a distance of $182^{+51}_{-33}$\,pc, which is about ten times closer than the DM distance estimates. Estimating pulsar distances using DM can be unreliable for sources away from the plane and PSR~\psrl has a Galactic latitude $b=-36.5\degr$. However, since data included in this study span only two years, we have have not included parallax in our final timing model fits and we hope to revisit this discrepancy with a longer timing baseline; the changes to other parameters when parallax is included are within their uncertainties published here. There are no GAIA counterparts within $5\arcmin$ of PSR~\psrl \citep{gaia_mission,gaia_DR3}.

\subsection{Black Widow Systems}
\begin{figure*}[t]
\centering
\includegraphics[width=0.9\textwidth]{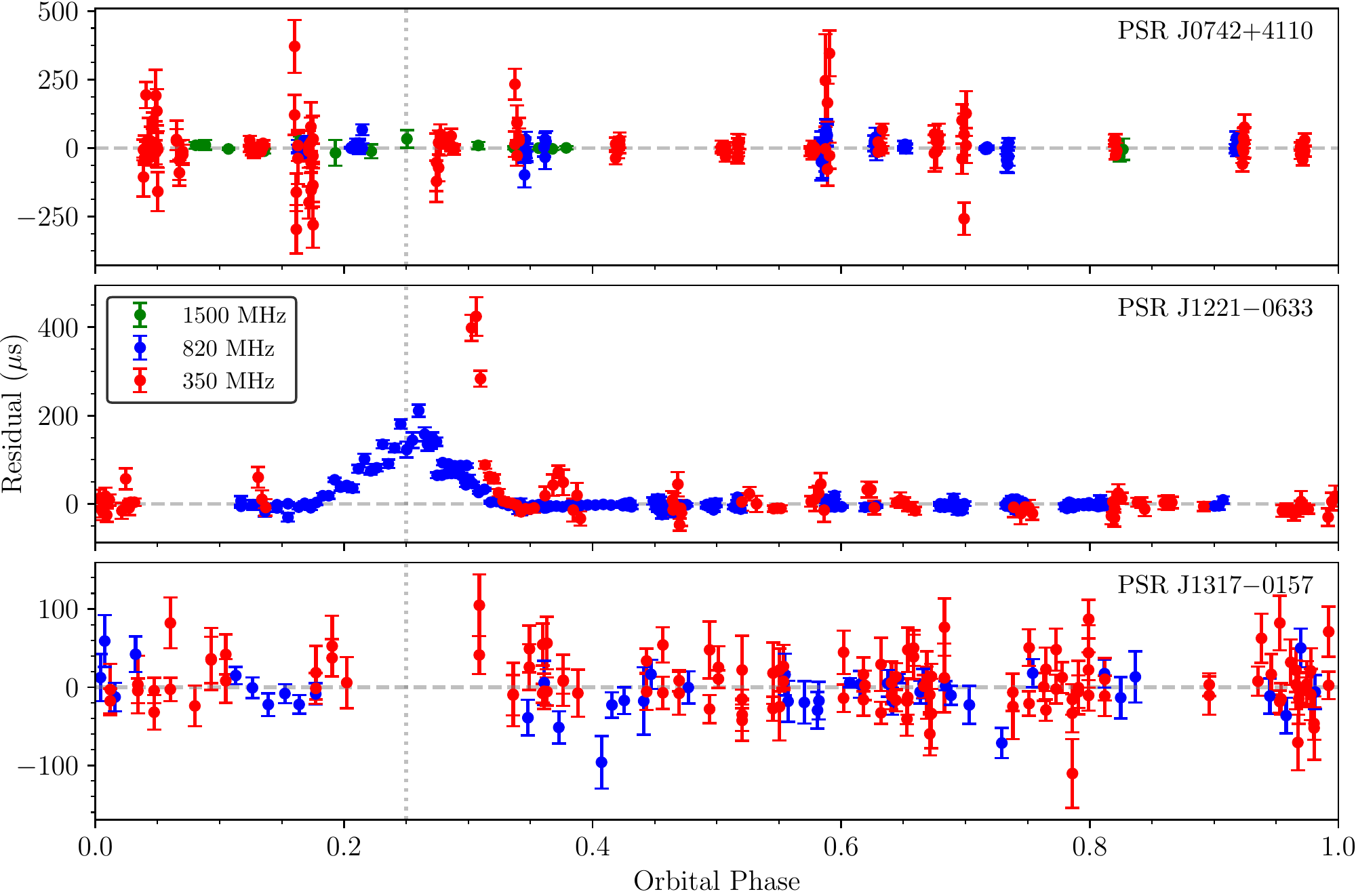}
\centering
\caption{Residuals from PSRs~\psrb, \psrf, and \psrg plotted against orbital phase. Superior conjunction at $\phi_{\rm orb}=0.25$ is shown with a dotted line. The lack of detections at or near superior conjunction indicates that the signal from PSR~\psrg is likely being eclipsed by its WD companion, and dispersive delays in PSR~\psrf's signal while behind its companion indicate partial eclipsing. Based on PSR~\psrb residuals, there appears to be no sign of eclipse.}
\label{fig:eclipse}
\end{figure*}

Both PSRs~\psrf and \psrg are in tight ($<10$\,hr) orbits, they likely have very low mass companions ($M_{\rm c} < 0.05$\,\Msun), and both exhibit eclipses (see Figure \ref{fig:eclipse}); all of these traits are consistent with ``black widow" systems \citep{fst+88,roberts+11} in which pulsars are actively ablating their companions. As a result, a large amount of intrabinary material is present, which can cause the pulsar signal to be additionally dispersed or completely obscured around superior conjunction. Sections \ref{sec:fermi-J1221} and \ref{sec:fermi-J1317} describe our findings that PSRs~\psrf and \psrg have \textit{Fermi}-LAT $\gamma$-ray counterparts, with the former also exhibiting $\gamma$-ray pulsations.

PSR~\psrf was confirmed at 350\,MHz as a new, nearby MSP, with a 1.9\,ms spin period and estimated DM distance of $\approx1$\,kpc. It has a relatively broad, single-component profile and faint signal at 350\,MHz, but exhibits much higher S/N and a complex, multi-component profile at 820\,MHz  (see Figure \ref{fig:profs}). Once a preliminary set of orbital parameters revealed PSR~\psrf's short orbital period ($P_{\rm B}=9.26$\,hours), a long, 2.5\,hour scan was scheduled spanning superior conjunction to check for signs of eclipsing. Figure \ref{fig:eclipse} shows the additional $\approx200$\,\us delay in pulse arrival times at 820\,MHz around superior conjunction, likely due to dispersion as the pulsar signal travels through plasma surrounding the companion (not Shapiro delay). This delay suggests an extra electron column density of $\approx10^{17}$\,cm$^{-2}$, which is comparable to similar measurements for other systems \citep{sbl+01,freire+05}. Although the duration of eclipses for pulsars in black widow systems is known to vary from one orbit to the next, our scan from MJD 58098 shows both ingress at $\phi_{\rm orb}\approx0.17$ and egress at $\phi_{\rm orb}\approx0.33$, so the pulsar signal is affected over $\approx16\%$ of an orbit (about 1.5\,hours).

PSR~\psrg is a 2.9\,ms pulsar in a 2.14\,hour orbit and like PSR~\psrf, it is also a new black widow system that shows signs of eclipses around superior conjunction. Unlike PSR~\psrf, we have not detected this pulsar's signal {\it during} eclipses, but on MJD 58116, PSR~\psrg was observed coming out of an eclipse during a 20\,minute scan. In this observation, the pulsar's signal is obscured for $\approx$\,500\,s, placing a lower limit on the duration of eclipse ($\approx$\,6.5\% of the orbit). The lack of detections over 10$-$15\% of the orbit (see Figure \ref{fig:eclipse}) suggests eclipses in this case tend to last for 13$-$20\,mins.

With estimates for the duration of eclipses for both PSRs~\psrf and \psrg, and assuming inclination angles of $90^{\circ}$, we place limits on each companion's radius, $R_{\rm c, min\, J1221}=1.2$\,\Rsun and $R_{\rm c, min\,J1317}=0.43$\,\Rsun. Also assuming edge-on orbits, and pulsar masses $m_{\rm p}=1.4$\,\Msun, orbital separations are $a_{J1221}=2.5$\,\Rsun and $a_{J1317}=0.94$\,\Rsun, and \citep[based on][]{egg+83}, the Roche lobe radius for each system is much smaller than the corresponding radius of each eclipsing object (by a factor of 5 in both cases). PSRs~\psrf and \psrg have companions with unbound plasma clouds, indicating that they are losing mass.

\subsection{PSR~\psrb}
PSR~\psrb is a 3.1\,ms pulsar in a short, 1.4\,day orbit around a low mass ($m_{\rm c,min}=0.06$\,\Msun) companion. Originally found in 2012, it was first published among the first batch of 67 GBNCC discoveries \citep{slr+14}, but initial timing follow-up was conducted using an incorrect position. As a result, deriving a coherent timing solution for this pulsar was significantly delayed. PSR~\psrb has a DM distance $<1$\,kpc (see Table \ref{tab:pms}), it appears to be relatively bright, and has a shallow spectrum (see Table \ref{tab:flux}), however, its profile is broad and lacks sharp features that might otherwise make it suitable for use in PTAs. Despite this, we measure significant proper motion for this pulsar (see Table \ref{tab:astrometric}), likely because of the comparatively longer timing baseline than other MSPs included here. The low companion mass for PSR~\psrb is near the threshold for those typical of black widow systems, however in combination with its orbital period, PSR~\psrb's properties are inconsistent with those of the black widow population. We also do not see any evidence of eclipsing or excess dispersion delays around superior conjunction (see Figure \ref{fig:eclipse}).

\begin{deluxetable}{lr}
  \tabletypesize{\footnotesize}
  \tablewidth{\columnwidth}
  \tablecaption{DD Binary Parameters of PSR J1018$-$1523}
  \tablecolumns{2}
  \tablehead{\colhead{Parameter}& \colhead{Value}}
  \startdata
  \cutinhead{Measured Parameters}
$P_{\rm B}$ (days)                         & 8.9839727(6) \\
$a\sin{i}/c$ (s)                           & 26.15662(3) \\
$T_0$ (MJD)                                   & 57545.74874(3) \\
$e$                                           & 0.227749(2) \\
$\omega$ ($\arcdeg$)                        & 60.013(1) \\
\cutinhead{Derived Parameters}
$f_{\rm M}$ (\Msun)     & 0.238062 \\
$M_{\rm c,min}$ (\Msun) & 1.16 \\
$\dot{\omega}$ ($\arcdeg$\,yr$^{-1}$) & 0.010(1) \\
$m_{\rm tot}$ (\Msun)    & 2.3(3) \\
  \enddata
  \tablecomments{Values in parentheses are the 1-$\sigma$
  uncertainty in the last digit as reported by \texttt{TEMPO}.}
\end{deluxetable}\label{tab:dns}

\begin{deluxetable*}{llllrrcc}
  \tabletypesize{\footnotesize}
  \tablewidth{\textwidth}
  \tablecaption{ELL1 Binary Parameters of GBNCC Pulsars}
  \tablecolumns{8}
  \tablehead{
  \colhead{PSR} & \multicolumn{5}{c}{Measured} & \multicolumn{2}{c}{Derived} \\
  \colhead{} &
  \colhead{$P_{\rm B}$ (days)} & \colhead{$a\sin{i}/c$ (s)} & 
  \colhead{$T_{\rm asc}$ (MJD)} &
  \colhead{$\epsilon_1$} & \colhead{$\epsilon_2$} &
  \colhead{$f_{\rm M}$ (\Msun)} & \colhead{$M_{\rm c,min}$ (\Msun)}}
  \startdata
  J0742+4110 & \phn1.385361182(2) & \phn0.556456(3) & 56045.146865(2) & 1(1)$\times10^{-5}$ & $-$0(9)$\times10^{-6}$ & 9.6394$\times10^{-5}$ & 0.06 \\
  J1045$-$0436 & 10.27364597(4) & 22.252633(8) & 57472.187456(2) & $-$4.53(8)$\times10^{-5}$ & 5.83(8)$\times10^{-5}$ & 1.1209$\times10^{-1}$ & 0.82 \\
  J1221$-$0633 & \phn0.386349620(4) & \phn0.0552855(7) & 57906.123003(2) & 1.0(3)$\times10^{-4}$ & $-$1.0(2)$\times10^{-4}$ & 1.2155$\times10^{-6}$ & 0.01 \\
  J1317$-$0157 & \phn0.089128297(2) & \phn0.027795(4) & 57909.041863(5) & 5(2)$\times10^{-4}$ & 0(2)$\times10^{-4}$ & 2.9024$\times10^{-6}$ & 0.02 \\
  J2022+2534 & \phn1.283702830(2) & \phn0.6092405(6) & 57535.5638685(8) & 2(2)$\times10^{-6}$ & 3(2)$\times10^{-6}$ & 1.4734$\times10^{-4}$ & 0.07 \\
  J2039$-$3616 & \phn5.789963674(4) & \phn3.3975854(4) & 57538.6033176(3) & $-$5.1(3)$\times10^{-6}$ & $-$3.3(3)$\times10^{-6}$ & 1.2562$\times10^{-3}$ & 0.14 \\
  \enddata
  \tablecomments{All timing models presented here use the ELL1 binary
  model, which is appropriate for low-eccentricity orbits.
  Values in parentheses are the $1$-$\sigma$ uncertainty in the last 
  digit as reported by \texttt{TEMPO}.}
  \end{deluxetable*}\label{tab:binary}

\subsection{PSR~\psrc}
PSR~\psrc is a recycled, 83\,ms pulsar in an eccentric ($e=0.23$), 9\,day orbit around a massive companion. Assuming a typical pulsar mass of 1.4\,\Msun, the minimum companion mass is $m_{\rm c,min}=1.16$\,\Msun. Based on the large companion mass and orbital eccentricity, PSR~\psrc is likely a new double neutron star (DNS) system \citep[for other examples see][]{tkf+17}. In addition to the five Keplerian parameters describing its orbit, we also find a significant change in the angle of periastron over time ($\dot{\omega}$; see Table \ref{tab:dns}), which is a relativistic effect predicted by GR. Compared to other known DNS systems \cite[summary provided in][]{tkf+17}, the Keplerian parameters here fall within typical ranges. The total mass derived from the advance of periastron ($\dot{\omega}$), $m_{\rm tot}=2.3\pm0.3$\,\Msun, is comparatively low, but it is consistent with other DNS systems to within 1-$\sigma$ uncertainty \citep{tkf+17,ligo_bns1,ligo_bns2}. Although PSR~\psrc is not expected to merge within a Hubble time, future work to improve the precision of its advance of periastron and possibly measure one or more additional post-Keplerian parameters will aid in constraining Galactic neutron star mass distributions \cite[see, e.g.,][]{fzt+19}.

\subsection{PSR~\psrd}
Although initially discovered as a bright candidate, five follow-up scans were needed to confirm PSR~\psrd and in future timing efforts, the pulsar was unreliably detected at both 350 and 820\,MHz. 
PSR~\psrd has a 24\,ms spin period and orbits an intermediate mass ($m_{\rm c,min}=0.82$\,\Msun) WD companion every 10.3\,days. With a nearly-circular orbit, PSR~\psrd's characteristics are similar to those of other intermediate mass binary pulsars \cite[IMBPs; e.g., see][]{clm+01,lorimer+08}, and most likely has a CO WD companion. There has been no correlation found between orbital phase and detectability, but because of its low DM (4.8\,\dmu), significant variability in the pulsar's apparent flux density due to scintillation is plausible. \cite{ymw+17} estimates PSR~\psrd has a scintillation timescale ($\Delta t=2,000$\,s) longer than a typical scan length and scintillation bandwidth ($\Delta f=130$\,MHz) comparable to the observing bandwidth at 820\,MHz. Assuming the DM distance is correct, PSR~\psrd has a height above the Galactic plane of $|z|\approx0.2$\,kpc, which is consistent with other IMBPs \citep{clm+01}.

\subsection{Optical Constraints}
For all of the binary pulsar systems in this study with nearly circular orbits (see Table \ref{tab:binary}), we have checked the  source catalogs of Pan-STARRS 3$\pi$ Steradian Survey \citep{cmm+16} for sources north of $\delta=-30\degr$ and SkyMapper \citep{skymapperDR2} for other sources  and find no optical/IR counterparts coincident with positions measured via pulsar timing.

Using the same procedure as that outlined in earlier studies \citep{lsk+18,kmk+18}, we use the average 5$\sigma$ magnitude lower limits for the PS1 {\tt griz} bands \citep[23.3, 23.2, 23.1, and 22.3 respectively;][]{cmm+16} to place constraints on PSR \psrd's CO-core WD companion (assuming an orbital inclination angle of 60$^\circ$ to calculate a median companion mass $M_{\rm c,med}=1.0$\,\Msun). We also estimate reddening along the line of sight based on a 3D map of interstellar dust reddening \citep{gsz+19} and the largest (most conservative) DM distance estimate available (see Table \ref{tab:derived}). Reddening values are converted to extinctions in PS1 bands using Table 6 in \cite{sf+11}. Comparing de-reddened magnitude limits to corresponding hydrogen-atmosphere (DA) cooling models\footnote{\url{http://www.astro.umontreal.ca/~bergeron/CoolingModels/}} from \cite{bwd+11}, we find the $i$-band limit to be the most constraining, limiting PSR \psrd's companion to an effective temperature $T_{\rm eff}<5200$\,K and age $>8.3$\,Gyr for the median companion mass.

PSR \psrl likely has a He-core WD companion based on its companion mass $M_{\rm c,med}=0.17$\,\Msun (see Table \ref{tab:binary}). Since this source is outside the region of sky coverered by Pan-STARRS, we use average 5$\sigma$ {\tt griz} limits from SkyMapper instead \citep[22, 22, 21, and 20, respecively;][]{skymapperDR2}. We use extremely-low mass (ELM) evolutionary models from \citet{althaus+13} together with hydrogen model atmospheres from \cite{bwd+11} to find that the $r$-band limit constrains the effective temperature to be $T_{\rm eff}<7800\,$K for an assumed radius of $0.04\,R_\odot$.  We verified that this radius is consistent with expectations for a $0.17\,M_\odot$ ELM WD based on custom-made evolutionary models computed following \cite{imt+16}.  Note that in this regime ELM WDs do not exhibit hydrogen shell flashes but cool steadily, allowing us to limit the age to $>9.8\,$Gyr since the end of Roche-lobe overflow.  However, at lower inclinations, $<45\degr$, or with a more massive pulsar the companion mass could exceed $0.2\,M_\odot$ and then the age constraint would not be useful.

Through pulsar timing, we find median masses for the remaining companions to be $\ll0.1$\,\Msun, which is below the range of expected WD masses \citep[e.g., see][]{imt+16}. Instead these are likely ``black widow" or ``redback" companions that are the remains of partially-degenerate stars ablated by the pulsars.  To constrain any possible companion, we first calculate Roche lobe radii for the companions based on \citet{egg+83}.  We then assume that the companions have a volumetric Roche lobe filling factor of 50\%.  Using the main sequence colors from \citet{cis+07} together with the main sequence radii from \citet{pm+13}, we infer effective temperature limits of $<2800\,$K for \psrb, \psrf, and \psrk\ (note that the limits for these sources could be more constraining, but this is the coolest effective temperature in the model grids).  For \psrg\ we have a less constraining limit of $T_{\rm eff}<4800\,$K.  Even for Roche lobe filling fractions of 10\% the limits on \psrb, \psrf, and \psrk\ stay at $<2800\,$K, which are consistent with typical night-side temperatures for known black widow systems \citep[e.g.,][]{breton+13,dhillon+22}.

We also examined archival data available through the Aladin server \footnote{\url{https://aladin.u-strasbg.fr/}} for associated diffuse structures such as pulsar wind nebulae and supernova remnants. Across all available wavelengths for each of the 12 pulsars, we searched for both catalogued objects and for symmetric diffuse emission reminiscent of a previously unidentified SNR or PWN. No plausible structures were found.

\subsection{Gamma-Ray Pulsars}
Two of our pulsars -- PSRs~J1221--0633 and J2039--3616 -- show detectable, pulsed gamma-ray emission (see Sections \ref{sec:fermi-J1221} and \ref{sec:fermi-J2039}). The gamma-ray efficiencies of 5--35\% are similar to those measured for other MSPs from which gamma-rays have been detected \citep{aaa+15}. As shown in Figure~\ref{fig:gray_profs}, the radio profile of  PSR~J1221--0633 has two components, with  the gamma-ray peak aligning with the weaker component. Conversely, the radio profile of PSR~J2039--3616 has only one component, and it is aligned with the gamma-ray profile. These properties appear to be broadly representative of those of MSPs published in the {\it Fermi} Second Pulsar Catalog, in which  27 MSPs had mis-aligned radio/gamma-ray profiles and six had aligned radio/gamma-ray profiles \citep{aaa+15}. Mis-aligned profiles are easily interpreted with  standard ``slot gap'' or ``outer gap'' emission models with narrow beams, while the aligned profiles require both the radio and gamma-ray emission to originate in the outer gap region. Aligned MSPs may be more likely to have low linear polarization due to caustic emission over a wide range of altitudes. Future polarization studies could test this hypothesis for PSR~J2039--3616. In addition, \citet{aaa+15}  found that pulsars with aligned radio/gamma-ray profiles generally had higher values of magnetic field at the light cylinder. PSR~J2039--3616 does not seem to fit this picture, however, as its magnetic field at the light cylinder is 4.4$\times10^4$~G and the mean value for the MSP population (i.e. periods greater than 30~ms) is 8.5$\times10^4$~G.

\section{Conclusion}\label{sec:conclusion}
We present coherent timing solutions for 12  pulsars discovered by the GBNCC pulsar survey. Seven of these discoveries are in binary systems: five MSPs orbiting low mass WD companions (including two black widow systems that eclipse and two high-precision timers suitable for pulsar timing arrays), one IMBP, and a new DNS system. Our results show that three discoveries (an isolated MSP and two DRPs) evolved via interaction with binary companions sometime in the past and two more are younger, isolated and non-recycled pulsars. These results underscore the importance of long-term pulsar timing; classifying evolutionary histories of systems like these requires spindown measurements, which can only be obtained with data spanning one year or more.

Since 2020, the GBNCC collaboration has been partnering with the CHIME Pulsar collaboration to extend existing pulsar timing solutions and rapidly follow up on new discoveries at Declination $\delta\gtrsim-15^{\circ}$. We are now regularly timing over 130 pulsars with CHIME, including several from this study, and those results will be presented in future work.

\section{Acknowledgements}
The Green Bank Observatory is a facility of the National Science Foundation (NSF) operated under cooperative agreement by Associated Universities, Inc. The National Radio Astronomy Observatory is a facility of the NSF operated under cooperative agreement by Associated Universities, Inc. J.K.S., D.L.K., M.A.M., M.E.D., T.D., S.M.R., and X.S. are supported by the NANOGrav NSF Physics Frontiers Center award numbers 1430284 and 2020265. E.P. is supported by an H2020 ERC Consolidator Grant `MAGNESIA' under grant agreement No. 817661 and National Spanish grant PGC2018-095512-BI00. Z.P. is a Dunlap Fellow. J.vL. acknowledges funding from the European Research Council under the European Union’s Seventh Framework Programme (FP/2007-2013) / ERC Grant Agreement n. 617199 (``ALERT"), and from Vici research programme ``ARGO" with project number 639.043.815, financed by the Netherlands Organisation for Scientific Research (NWO). M.A.M. and E.F.L. are supported by NSF OIA-1458952 and NSF Award Number 2009425. S.M.R. is a CIFAR Fellow. Pulsar research at UBC is supported by an NSERC Discovery Grant and by the Canadian Institute for Advanced Research. M.S. acknowledges funding from the European Research Council (ERC) under the European Union’s Horizon 2020 research and innovation programme (grant agreement No. 694745). M.E.D. acknowledges support from the Naval Research Laboratory by NASA under contract S-15633Y. This work has made use of data from the European Space Agency (ESA) mission {\it Gaia} (\url{https://www.cosmos.esa.int/gaia}), processed by the {\it Gaia}
Data Processing and Analysis Consortium (DPAC, \url{https://www.cosmos.esa.int/web/gaia/dpac/consortium}). Funding for the DPAC has been provided by national institutions, in particular the institutions participating in the {\it Gaia} Multilateral Agreement. T.D. is supported by an NSF Astronomy and Astrophysics Grant (AAG) award number 2009468.


\facility{GBT (GUPPI)}
\software{{\tt PRESTO} (\url{https://www.cv.nrao.edu/~sransom/presto/}),
	{\tt PSRCHIVE} \citep{hsm+04}, 
	{\tt TEMPO} (\url{http://tempo.sourceforge.net/)},
	{\tt PyGDSM} (\url{https://github.com/telegraphic/pygdsm}),
	{\tt PINT} (\url{https://github.com/nanograv/PINT}),
	{\tt PyGEDM} (\url{https://github.com/FRBs/pygedm})}

\end{document}